\documentclass[10pt]{article}
\usepackage{epsfig,amsmath,graphicx,amssymb,overpic,cite,amsthm}

\def\be{\begin{equation}}
\def\ee{\end{equation}}
\def\bee{\begin{eqnarray}}
\def\ene{\end{eqnarray}}
\def\bes{\begin{subequations}}
\def\ees{\end{subequations}}
\def\no{\nonumber}
\def\Re{{\rm Re}\,}
\def\Im{{\rm Im}\,}

\def\v{\vspace{0.1in}}
\def\d{\displaystyle}
\def\l{\left}
\def\r{\right}

\begin{document}

\baselineskip=13pt
\renewcommand {\thefootnote}{\dag}
\renewcommand {\thefootnote}{\ddag}
\renewcommand {\thefootnote}{ }

\pagestyle{plain}

\begin{center}
\baselineskip=16pt \leftline{} \vspace{-.3in} {\Large \bf Long-time asymptotics for initial-boundary value problems of integrable Fokas-Lenells equation
on the half-line} \\[0.2in]
\end{center}

\begin{center}
Shuyan Chen$^{a,b}$, Zhenya Yan$^{a,b,*}$\footnote{$^{*}${\it Email address}: zyyan@mmrc.iss.ac.cn}  \\[0.1in]
{\it \small $^{a}$Key Laboratory of Mathematics Mechanization, Academy of Mathematics and Systems Science, Chinese Academy of Sciences, Beijing 100190, China \\
    $^{b}$School of Mathematical Sciences, University of Chinese Academy of Sciences, Beijing 100049, China}
\end{center}

\vspace{0.1in}

{\baselineskip=15pt

\begin{tabular}{p{16cm}}
 \hline \\
\end{tabular}

\vspace{-0.18in}

\begin{abstract} \small \baselineskip=14pt
We study the Schwartz class of initial-boundary value (IBV) problems for the integrable Fokas-Lenells equation on the half-line via the Deift-Zhou's nonlinear descent method analysis of the corresponding Riemann-Hilbert problem such that the asymptotics of the Schwartz class of IBV problems as $t\to\infty$ is presented.

\vspace{0.1in} \noindent {\it MSC:}  37K40; 35Q55; 35Q15

\vspace{0.1in} \noindent {\it Keywords:}  \ Riemann-Hilbert problem; Fokas-Lenells equation; Initial-boundary value problem; Schwartz class; Nonlinear steepest descent method; Long-time asymptotics

\end{abstract}

\vspace{-0.05in}
\begin{tabular}{p{16cm}}
  \hline \\
\end{tabular}


\section{Introduction}

It is of important significance to explore the basic properties of the integrable nonlinear evolution equations with Lax pairs.   The completely integrable Fokas-Lenells (FL) equation~\cite{focas1,lenells1}
\bee  \label{yuanshi}
iq_{t}-\alpha q_{tx}+\gamma q_{xx}+\sigma|q|^2(q+i\alpha q_{x})=0, \quad i=\sqrt{-1}
\ene
is associated with the well-known nonlinear Schr\"odinger (NLS) equation, where $q(x,t)$ is a complex-valued function, the subscript denotes the partial derivative, $\alpha$ and $\gamma$ are real constants, $\sigma=\pm 1$. Similarly to the NLS equation~\cite{op}, the FL equation can also be derived from the Maxwell's equations and describes nonlinear pulse propagation in monomode optical fibers in the presence of higher-order nonlinear effects~\cite{lsam}. The bi-Hamiltonian structures, Lax pair, solitons, and the initial value problem of Eq.~(\ref{yuanshi}) were studied~\cite{lenells1}. The $N$-bright soliton~\cite{nbs} and dark soliton~\cite{nds} solutions of Eq.~(\ref{yuanshi}) were obtained via
the dressing method and B\"acklund transformation, respectively. Eq.~(\ref{yuanshi}) with $\alpha=0$ reduces to the NLS equation. Eq.~(\ref{yuanshi}) is associated with a variational principle $ iq_t=\frac{\delta H}{\delta \bar q}$ with the Hamiltonian being
 \bee
 H=\int_{R}\l[-\alpha |q_t|^2+\gamma |q_x|^2+\frac{\sigma}{2}(|q|^4+i\alpha |q|^2\bar qq_x)\r]dx.
 \ene

 Replacing $q(x,t)$ by $q(-x,t)$ and assuming that $\alpha \gamma>0$, one can use the gauge transformation $q(x,t)\rightarrow \sqrt{\frac{\gamma}{\alpha^3}}e^{\frac{ix}{\alpha}}q(x,t)$ and $\sigma \rightarrow -\sigma$
to change Eq.~(\ref{yuanshi}) into~\cite{lenells2}
\bee  \label{yuanshi2}
q_{tx}+\frac{\gamma}{\alpha^3}q-\frac{2i\gamma}{\alpha^2}q_{x}-\frac{\gamma}{\alpha}q_{xx}+i\sigma \frac{\gamma}{\alpha^3}|q|^2q_{x}=0 , \ \ \sigma=\pm 1.
\ene
 The initial-boundary value problem (IBVP) of Eq.~(\ref{yuanshi2}) with $\gamma=\alpha=\sigma=1$ formulated on the half-line was investigated~\cite{lenells2} through the Fokas' method~\cite{fokas}. In particular, the so-called linearizable boundary conditions were used to find explicit expressions for the spectral functions based on the inverse scattering method. Recently, the long-time asymptotics of Eq.~(\ref{yuanshi2}) with decaying initial-value problem (IVP) on the full-line was studied~\cite{xu15} by employing
 the Deift-Zhou's nonlinear descent method analysis~\cite{dz} of the Riemann-Hilbert problem (RHP) found in Ref.~\cite{lenells1}.

The Deift-Zhou's nonlinear steepest descent method~\cite{dz,dzb,dzc} has been used to study the long-time asymptotic behaviors of solutions of
IVPs of some nonlinear integrable systems based on the analysis of the corresponding RHPs.
After that, the Fokas' unified method~\cite{fokas} was presented to construct the matrix RHPs for the IBVPs of some linear and nonlinear integrable systems (see, e.g, the book~\cite{fokasbook} and references therein). Particularly, the two kinds of above-mentioned powerful methods have been effectively combined to investigate the long-time asymptotics of solutions of IBVPs for some nonlinear integrable equations such as the NLS equation, mKdV equation, sine-Gordon equation,  Degasperis-Procesi equation, derivative NLS equation, and KdV equation (see, e.g.,~\cite{ibv1,ibv2,ibv3,ibv4,ibv5,ibv6,ibv7,ibv8,ibv9}).

The aim of this paper is to explore the long-time asymptotics of the solution on the
half-line of the IBVP of Eq.~(\ref{yuanshi2}) with $\gamma=\alpha=\sigma=1$ as
\bee
\label{IBV}
\begin{array}{l}
q_{tx}+q-2iq_{x}-q_{xx}+i|q|^2q_{x}=0 \quad (\rm FL \,\, equation), \v\\
                      q(x,0)=q_{0}(x)\in S(\mathbb{R^+}) \quad ({\rm initial \,\, value \,\, condition}), \v\\
                    q(0,t)=g_{0}(t)\in S(\mathbb{R^+}) \quad ({\rm Dirichlet\,\, boundary \,\, value \,\, condition}), \v\\ q_x(0,t)=g_{1}(t)\in S(\mathbb{R^+}) \quad ({\rm Neumann \,\,boundary \,\, value \,\, conditions}),
\end{array}
 \ene
where $\mathbb{R^+}=[0, \infty)$, the Schwartz class is defined by
\bee\label{s}
 S(\mathbb{R^+})=\left\{f(y)\in \mathbb{C}^{\infty}(\mathbb{R^+})|y^\iota f^{(\beta)}(y)\in L^{\infty}(\mathbb{R^+}),\, \iota, \beta\in \mathbb{Z}_+\right\}.
 \ene

We will follow  the approach of \cite{ibv8, ibv9} and use the RH problem~\cite{lenells2} to
explore the long-time asymptotics of the Schwartz class of IBV problem
of the FL equation (\ref{IBV}) on the half-line using the nonlinear descent method. The main results of this paper are summarized in the following Theorem.\\

\noindent {\bf Theorem 1.1} {\it  Suppose that the initial-boundary values $q(x,0)=q_{0}(x),\, q(0,t)=g_{0}(t),\, q_x(0,t)=g_{1}(t)$ belong to
the Schwartz class (\ref{s}), the reflection coefficient $r(\lambda)$ defined by Eq.~(\ref{rm}) is determined via the spectral functions $a(\lambda),\, b(\lambda),\, A(\lambda), B(\lambda)$ defined by Eqs.~(\ref{sf}) related to the initial-boundary values, and Assumption 2.1 holds. Let $q(x,t)$ be the half-line solution of the initial-boundary values of the FL equation given by  Eq.~(\ref{IBV}). Then  for $0<x/t<N$ with $x>0,\, N>0$,  when $t\to \infty$, we know that  $q_x(x,t)$ has the long-time asymptotic as
\bee
q_x(x,t)=\frac{1}{\sqrt{x+t}}\left[\sqrt{\upsilon}e^{i[\eta_1(\lambda_0)
+\tau(t)]}-\sqrt{\tilde{\upsilon}}e^{i[\eta_2(\lambda_0)+\tau(t)]}\right]+O\l(\frac{\ln t}{t}\r), \,\,\,\, x, t>0,
\ene
where $\lambda_0=\sqrt[4]{\frac{1}{4(x/t+1)}}$,
\bee\no\begin{array}{c}
\upsilon=-\d\frac{1}{2\pi}\ln(1-|r(\lambda_0)|^2)\geq 0, \v\\
\tilde{\upsilon}=-\d\frac{1}{2\pi}\ln(1+|r(i\lambda_0)|^2)\leq 0,
\end{array}
\ene
the phases are given by
\bee\no\begin{array}{c}
\tau(t)=\d\int_0^t(|g_1(t')|^2-|g_0(t')|^2)dt'-4\int_0^{\lambda_0}\l[\frac{\upsilon(\lambda')+|\tilde{\upsilon}(\lambda')|}{\lambda'}
+\frac{\sqrt{\upsilon|\tilde{\upsilon}|(\lambda')}}{\lambda'}\sin(\eta_2(\lambda')-\eta_1(\lambda'))\r]d\lambda',\v\\
\eta_1(\lambda)|_{\lambda=\lambda_0}=\d\frac{\pi}{4}+\arg r(\lambda_0)-\arg\Gamma(-i\upsilon(r(\lambda_0))) +
2\tilde{\upsilon}\ln{2\lambda_0^2}-\upsilon\ln\frac{\lambda_0^2}{t} \v\\
+(2-\lambda_0^{-2})t+2i[\chi_{\pm}(\lambda_0)+\widetilde{\chi'}_{\pm}(\lambda_0)],\v\\
\eta_2(\lambda)|_{\lambda=\lambda_0}=\d\frac{\pi}{4}+\arg r(i\lambda_0)+\arg\Gamma(i\tilde{\upsilon}(r(i\lambda_0)))-
2\upsilon\ln{2\lambda_0^2}+\tilde{\upsilon}\ln\frac{\lambda_0^2}{t} \v\\
+(2+\lambda_0^{-2})t +2i[\chi'_{\pm}(i\lambda_0)+\widetilde{\chi}_{\pm}(i\lambda_0)]
\end{array}
\ene
with
\bee\no\begin{array}{c}
\chi_{\pm}(\lambda)=\d\frac{1}{2\pi i}\int_{0}^{\pm\lambda_0} \ln \left(\frac{1-|r(\lambda')|^2}{1-|r(\lambda_0)|^2} \right) \frac{\mathrm{d}\lambda'}{\lambda'-\lambda},\v\\
\tilde{\chi}_{\pm}(\lambda)=\d\frac{1}{2\pi i}\int_{\pm i\lambda_0}^{i0} \ln \left(\frac{1-r(\lambda')\overline{r(\overline{\lambda'})}}{1+|r(i\lambda_0)|^2} \right) \frac{\mathrm{d}\lambda'}{\lambda'-\lambda},\v\\
\chi'_{\pm}(z)=\d \exp\l[\frac{i}{2\pi}\int_{0}^{\pm \lambda_0} \ln |z-z'|d\ln(1-|r(z')|^2)\r],\v\\
\tilde{\chi}'_{\pm}(z)=\d \exp\l[\frac{i}{2\pi}\int_{\pm \lambda_0}^{0} \ln |z-iz'|d\ln(1+|r(iz')|^2)\r].
\end{array}
\ene
and $\Gamma$ denoting the Gamma function.}

In the following several sections, we would like to proof Theorem 1.1.

\section{Preliminaries}

{\it 2.1. \, Lax pair} \v

Eq.~(\ref{IBV}) possesses the following Lax pair~\cite{lenells2}
\begin{eqnarray}
    \begin{array}{l}
       \psi_x=W\psi,\quad W=-i\lambda^2\sigma_{3}+\lambda U_x,    \vspace{0.1in}  \\
        \psi_{t}=V\psi,\quad V=-i\eta^2\sigma_{3}+\lambda U_x-\frac{i}{2}\sigma_3U^2+\frac{i}{2\lambda}\sigma_3U,
                 \end{array}
                 \label{yuanlaxpair}
\end{eqnarray}
where $\psi=\psi(x,t)$ is a 2$\times$2 matrix-valued eigenfunction, $\lambda\in \mathbb{C}$ is an isospectral parameter, and
\bee\no
U= \left(
                 \begin{array}{cc}
                 0       &q(x,t) \v\\
                    \bar{q}(x,t)             & 0
                 \end{array}
                 \right),
                 \ \ \ \ \ \
\sigma_{3}=\left(
                 \begin{array}{cc}
                 1  &0 \v\\    0                 & -1
                 \end{array}
                 \right), \ \ \ \ \ \
\eta=\lambda-\frac{1}{2\lambda},
\ene
where $\bar{q}$ denotes the complex conjugate of the potential function $q$. It is easy to see that the compatible conditions $\psi_{xt}=\psi_{tx}$, that is, the zero-curvature equation
$W_t-V_x+[W, V]=0$, just generates the FL equation (\ref{IBV}).

To conveniently solve the eigenfunction, we use the transformation
\bee
\d\psi(x,t,\lambda)=e^{i\int_{(0,0)}^{(x,t)}\Delta \sigma_3}\mu(x,t,\lambda)
e^{-i\int_{(0,0)}^{(\infty,0)}\Delta \sigma_3}
e^{-i\theta\sigma_3},\quad \theta=\lambda^2x+\eta^2t,
\ene
where $\Delta$ is given by the closed real-valued one-form
\bee
\Delta(x,t)=\frac{1}{2}|q_x|^2dx+\frac{1}{2}(|q_x|^2-|q|^2)dt.
\ene
 \vspace{0.05in}
Then the function $\mu$ satisfies the following Lax pair
\bee \label{laxmu}
 \begin{array}{l}
 \mu_x+i\lambda^2[\sigma_3, \mu]=V_1\mu, \vspace{0.1in} \\
\mu_t+i\eta^2[\sigma_3, \mu]=V_2\mu,
\end{array}
\ene
where the matrices $V_1$ and $V_2$ are given by
\bee\no\begin{array}{l}
V_1= \left(    \begin{array}{cc}
                 -\dfrac{i}{2}|q_x|^2       &\lambda q_xe^{-2i\int_{(0,0)}^{(x,t)}\Delta} \vspace{0.1in}\\
                    \lambda \bar{q}_xe^{2i\int_{(0,0)}^{(x,t)}\Delta}            & \dfrac{i}{2}|q_x|^2
                 \end{array}
                 \right), \v\\
V_2= \left(
                 \begin{array}{cc}
                 -\dfrac{i}{2}|q_x|^2       & \l(\lambda q_x+\dfrac{i}{2\lambda}q\r)e^{-2i\int_{(0,0)}^{(x,t)}\Delta} \vspace{0.1in}\\
                    \l(\lambda \bar{q}_x-\dfrac{i}{2\lambda}\bar{q}\r)e^{2i\int_{(0,0)}^{(x,t)}\Delta}            & \frac{i}{2}|q_x|^2
                 \end{array}
                 \right).
\end{array}
\ene

\v
\noindent {\it 2.2.\,  Riemann-Hilbert problem for the FL equation with the IBV conditions} \v

Suppose that the  initial data $q(x,0)=q_0(x)$, the Dirichlet and Neumann boundary values $q(0,t)=g_0(t)$ and $q_x(0,t)=g_1(t)$
belong to the Schwartz class $S(\mathbb{R^+})$. To express the solution of Eq.~(\ref{IBV}) on the half-line by means of
the solution of a $2\times 2$ matrix-valued RH problem, we define the four spectral functions  \{$a(\lambda),b(\lambda),A(\lambda),B(\lambda)$\} by~\cite{lenells2}
\bee\label{sf}
X(0,\lambda)= \left(
                 \begin{array}{cc}
                   \overline{ a(\overline{\lambda})} &b(\lambda) \vspace{0.1in}\\
                            \overline{ b(\overline{\lambda})}   & a(\lambda)
                 \end{array}
                 \right), \  \  \
T(0,\lambda)= \left(
                 \begin{array}{cc}
                   \overline{ A(\overline{\lambda})} &B(\lambda) \vspace{0.1in}\\
                            \overline{ B(\overline{\lambda})}   & A(\lambda)
                 \end{array}
                 \right),
\ene
where $X(x,\lambda)$ and $T(t,\lambda)$ are defined by the Volterra integral equations
\bee\begin{array}{l}
X(x,\lambda)=I+\d\int_\infty^x e^{i\lambda^2(\xi-x)\hat{\sigma}_3}V_1(\xi,0,\lambda)X(\xi,\lambda)d\xi, \v\\
T(t,\lambda)=I+\d\int_\infty^t e^{i\eta^2(\tau-t)\hat{\sigma}_3}V_2(0,\tau,\lambda)T(\tau,\lambda)d\tau,
\end{array}
\ene
where the operator $e^{\hat{\sigma}_{3}}$ acts on a $2\times 2$ matrix $A$ by $e^{\hat{\sigma}_{3}}A=e^{\sigma_3} Ae^{-\sigma_3}$.

To study the properties of \{$a(\lambda),b(\lambda),A(\lambda),B(\lambda)$\}, we define the following the open domains of the complex $\lambda$-plane by (see Fig.~\ref{fig1})~\cite{lenells2}
\bee
\begin{array}{l}
D_1=\l\{\lambda\in \mathbb{C}| {\rm arg}\lambda\in (0,\frac{\pi}{2})\cup (\pi,\frac{3\pi}{2})  \ \ {\rm and}  \ \ |\lambda|>\frac{\sqrt{2}}{2}\r\}, \vspace{0.1in}\\
D_2=\l\{\lambda\in \mathbb{C}| {\rm arg}\lambda\in (0,\frac{\pi}{2})\cup (\pi,\frac{3\pi}{2})  \ \ {\rm and}  \ \ |\lambda|<\frac{\sqrt{2}}{2}\r\}, \vspace{0.1in}\\
D_3=\l\{\lambda\in \mathbb{C}| {\rm arg}\lambda\in (\frac{\pi}{2},\pi)\cup (\frac{3\pi}{2},2\pi)  \ \ {\rm and}  \ \ |\lambda|<\frac{\sqrt{2}}{2}\r\}, \vspace{0.1in}\\
D_4=\l\{\lambda\in \mathbb{C}| {\rm arg}\lambda\in (\frac{\pi}{2},\pi)\cup (\frac{3\pi}{2},2\pi)  \ \ {\rm and}  \ \ |\lambda|>\frac{\sqrt{2}}{2}\r\}.
\end{array}
\ene

\begin{figure}[!t]
\begin{center}
\vspace{0.05in}
{\scalebox{0.4}[0.3]{\includegraphics{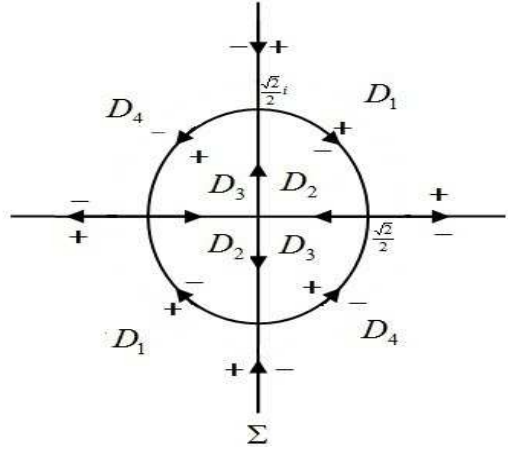}}}
\end{center}
\vspace{-0.2in}\caption{The original contour $\Sigma$ for $M(x,t,k)$. }
\label{fig1}
\end{figure}

Thus we have the properties of the spectral functions  \{$a(\lambda),b(\lambda),A(\lambda),B(\lambda)$\} (cf. Ref.~\cite{lenells2}):
\begin{itemize}
\item{} $a(\lambda)$ and $b(\lambda)$ are continuous and bounded for $ \overline{D}_1\cup \overline{D}_2$ and analytic in $D_1\cup D_2$;
\item{} $a(\lambda)\overline{a(\bar{\lambda})}-b(\lambda)\overline{b(\bar{\lambda})}=1,\, \lambda \in \overline{D}_1\cup \overline{D}_2$;
\item{} $a(\lambda)=1+O(\frac{1}{\lambda})$ and $b(\lambda)=\frac{b_1}{\lambda}+O(\frac{1}{\lambda^2})$ uniformly when $\lambda\rightarrow\infty,\lambda\in D_1\cup D_2;$
\item{} $A(\lambda)$ and $B(\lambda)$ are continuous and bounded for $ \overline{D}_1\cup \overline{D}_3$ and analytic in $D_1\cup D_3$;
\item{} $A(\lambda)\overline{A(\bar{\lambda})}-B(\lambda)\overline{B(\bar{\lambda})}=1,\lambda \in \overline{D}_1\cup \overline{D}_3$;
\item{} $A(\lambda)=1+O(\frac{1}{\lambda})$ and $B(\lambda)=\frac{B_1}{\lambda}+O(\frac{1}{\lambda^2})$ uniformly when $\lambda\rightarrow\infty,\, \lambda\in D_1\cup D_3.$
\end{itemize}

To present the RH problem of Eq.~(\ref{IBV}), we give the following assumption:\v

\noindent \textbf{Assumption \, 2.1}\, We assume that the spectral functions  \{$a(\lambda),b(\lambda),A(\lambda),B(\lambda)$\} and
initial-boundary values $\{q_0(x),\, g_0(t),\, g_1(t)\}$ satisfy the following conditions~\cite{lenells2}:
\begin{itemize}
\item[i)] The above-defined spectral functions \{$a(\lambda),b(\lambda),A(\lambda),B(\lambda)$\}  satisfy the global relation
$a(\lambda)B(\lambda)-b(\lambda)A(\lambda)=0$;

\item[ii)] $a(\lambda)$ and $d(\lambda)=a(\lambda)\overline{A(\bar{\lambda})}-b(\lambda)\overline{B(\bar{\lambda})}$ have no zeros in $\overline{D}_1\cup \overline{D}_2$ and $\overline{D}_1$, respectively;

\item[iii)] initial-boundary value conditions $q(x,0)=q_{0}(x)$, $q(0,t)=g_{0}(t)$, and $q_x(0,t)=g_{1}(t)$ are compatible for  Eq.~(\ref{IBV}) to all orders at $x=t=0$, that is,
\bee\no
q_{0}(0)=g_{0}(0),\quad  g_{1}(0)=q_{0x}(0),\quad g_{1t}(0)+q_{0}(0)-2iq_{0x}(0)-q_{0xx}(0)+i|q_{0}(0)|^2g_{1}(0)=0.
\ene
\end{itemize}
then we call $\{g_{0}(t),g_{1}(t)\}$ are admissible set of functions with respect to $q_{0}(x)$.

Similarly to Ref.~\cite{lenells2}, if Assumption 2.1 is satisfied, then a RH problem related to Eq.~(\ref{IBV}) given by
\bee\label{orh}
\left\{\begin{array}{l}
 M(x,t,\lambda) {\rm \,\,is \,\, in\,\, general\,\, a\,\, meromorphic \,\,function\,\, in\,\,} \lambda\in\mathbb{C}\backslash\Sigma, \v\\
 M_{+}(x,t,\lambda)=M_{-}(x,t,\lambda)J (x,t,\lambda) {\rm \,\,for\,\,} \lambda\in \overline{D}_i\cap \overline{D}_j,\,\, i,j=1,2,3,4, \v\\
M(x,t,\lambda)=I+O(\frac{1}{\lambda}) {\rm\,\, as\,\,} \lambda\rightarrow\infty.
\end{array}
\right.
\ene
has a unique solution $M(x,t,\lambda)$ for $(x,t)\in \mathbb{R}^+\times \mathbb{R}^+$, where the jump matrix $J(x,t,\lambda)$ is defined by
\bee
J(x,t,\lambda)=\left\{
                 \begin{array}{ll}
          J_1, & \lambda \in \overline{D}_1\cap \overline{D}_2, \vspace{0.1in}  \\
          J_2=J_1J^{-1}_4J_3, & \lambda \in \overline{D}_2\cap \overline{D}_3, \vspace{0.1in}  \\
          J_3, & \lambda \in \overline{D}_3\cap \overline{D}_4, \vspace{0.1in}  \\
          J_4, & \lambda \in \overline{D}_4\cap \overline{D}_1,
                 \end{array}
                 \right.
\ene
with
\begin{center}
$J_1=\left(\begin{array}{cc}  1& 0 \v\\   -\Omega(\lambda)e^{2i\theta}& 1
                 \end{array}
                 \right), \quad
J_3=\left(
                 \begin{array}{cc}
           1&  \overline{\Omega(\bar{\lambda})}e^{-2i\theta} \v\\
               0& 1                 \end{array}
                 \right),\quad
           J_4=\left(
                 \begin{array}{cc}
            \dfrac{1}{a(\lambda)\overline{a(\bar{\lambda})}}& \dfrac{b(\lambda)}{\overline{a(\bar{\lambda})}}e^{-2i\theta} \v\\
               -\dfrac{\overline{b(\bar{\lambda})}}{a(\lambda)}e^{2i\theta}&1
                 \end{array}
                 \right), $
\end{center}
and
\bee
\Omega(\lambda)=\frac{\overline{B(\bar{\lambda})}}{a(\lambda)d(\lambda)},\quad d(\lambda)=a(\lambda)\overline{A(\bar{\lambda})}-b(\lambda)\overline{B(\bar{\lambda})},\quad
 \lambda\in \overline{D}_2.
\ene

Thus we have the solution of the FL equation with the initial-boundary values (\ref{IBV}) in the form(cf.~\cite{lenells2})
\bee
q(x,t)=-2i\int_x^\infty m(\xi,t)e^{2i\int_{(0,0)}^{(\xi,t)}\Delta}d\xi,
\ene
where
\bee
m(x,t)=\lim\limits_{\lambda\rightarrow\infty}(\lambda M(x,t,\lambda))_{12},\quad
\Delta=2|m|^2dx-2\left(\int_x^\infty (|m(\xi,t)|^2)_t d\xi\right)dt,
\ene
with $M(x,t,\lambda)$ being defined by the RH problem (\ref{orh}).

\begin{figure}[!t]
\begin{center}
{\scalebox{0.3}[0.3]{\includegraphics{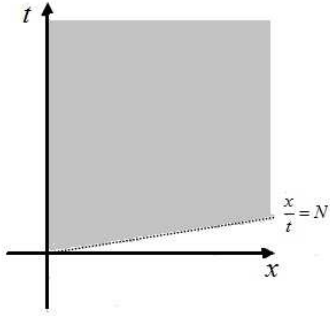}}}
\end{center}
\vspace{-0.8in}\caption{The asymptotic sector $0<\frac{x}{t}<N$ (shaded). }
\end{figure}

Our main result presents an explicit formula for the long-time asymptotics of the solution $q(x,t)$ of the FL equation on the half-line under the IBV lied in the Schwartz class. The result of this paper is valid in the sector $0<\frac{x}{t}<N$ exhibited in Figure 2. Moreover,
the solution of the FL equation on the sector $N<\frac{x}{t}<\infty$ is seen as the absence of boundaries, and has been investigated~\cite{xu15}.

\section{Modifications of the original RH problem}

\v
\noindent {\it 3.1.\, The first modification (modified reflection coefficients)}
\v

For the half-line problem, the coefficients typically only decay like $\frac{1}{\lambda}$ as $\lambda\rightarrow\infty$. Therefore, we transform the matrix $M(x,t,\lambda)$ in the original RH problem (\ref{orh}) by introducing the sectionally analytic function $M^{(1)}(x,t,\lambda)$ by
\bee
M(x,t,\lambda)=M^{(1)}(x,t,\lambda)F^{(1)}(x,t,\lambda),
\ene
where the transformation $F^{(1)}(x,t,\lambda)$ is given by
\begin{center}
$F^{(1)}(x,t,\lambda)=\left\{
                 \begin{array}{ll}
          \left(
                 \begin{array}{cc}         1& 0 \v\\
               -\d\frac{\lambda \bar{B}_1}{\lambda^2+1}e^{t\Phi}& 1\\
                 \end{array}
                 \right), & \lambda \in D_1, \vspace{0.1in}  \\
        \left(
                 \begin{array}{cc}
           1& -\d\frac{\lambda B_1}{\lambda^2+1}e^{-t\Phi} \v\\
               0& 1\\
                 \end{array}
                 \right),  & \lambda \in D_4, \vspace{0.1in}  \\
          I,& {\rm elsewhere}.
                 \end{array}
                 \right.$
\end{center}
with
$\Phi=2i(\frac{x}{t}\lambda^2+\eta^2)$. The factor $\frac{\lambda \bar{B}_1}{\lambda^2+1}$ is an odd function of $\lambda$ which is analytical in
$D_1$ (the poles lie at $\lambda=\pm e^{i\pi}$) such that
\bee
\frac{\lambda \bar{B}_1}{\lambda^2+1}=\frac{\bar{B}_1}{\lambda}+O(\lambda^{-2}),\quad \lambda \to\infty.
\ene

Therefore, we know that $M(x,t,\lambda)$ satisfies the original RH problem (\ref{orh}) if and only if $M^{(1)}(x,t,\lambda)$ solves the following first-modification RH problem
\bee\label{orh1}
\left\{\begin{array}{l}
 M^{(1)}(x,t,\lambda) {\rm \,\,is \,\, in\,\, general\,\, a\,\, meromorphic \,\,function\,\, in\,\,} \lambda\in\mathbb{C}\backslash\Sigma, \v\\
 M_{+}^{(1)}(x,t,\lambda)=M_{-}^{(1)}(x,t,\lambda)J^{(1)} (x,t,\lambda) {\rm \,\,for\,\,} \lambda\in \overline{D}_i\cap \overline{D}_j,\,\, i,j=1,2,3,4, \v\\
M^{(1)}(x,t,\lambda)=I+O(\frac{1}{\lambda}) {\rm\,\, as\,\,} \lambda\rightarrow\infty.
\end{array}
\right.
\ene
where the jump matrix $J^{(1)}=F^{(1)}_-J (F^{(1)})^{-1}_+$ defined by
\begin{center}
$J^{(1)}(x,t,\lambda)=\left\{
                 \begin{array}{ll}
          \left(
                 \begin{array}{cc}
           1& 0 \v\\
               \left(\dfrac{\lambda \bar{B}_1}{\lambda^2+1}-\Omega(\lambda)\right)e^{t\Phi}& 1
                 \end{array}
                 \right), & \lambda \in \overline{D_1}\cap \overline{D_2}, \vspace{0.1in}  \\
           \left(
                 \begin{array}{cc}
           1& 0 \v\\
               \l(\dfrac{\overline{b(\bar{\lambda})}}{a(\lambda)}-\Omega(\lambda)\r)e^{t\Phi}&1
                 \end{array}
                 \right)
                 \left(
                 \begin{array}{cc}
           1& \l(\overline{\Omega(\bar{\lambda})}-\dfrac{b(\lambda)}{\overline{a(\bar{\lambda})}}\r)e^{-t\Phi} \v\\
               0&1
                 \end{array}
                 \right),
                 & \lambda \in \overline{D_2}\cap \overline{D_3}, \vspace{0.1in}  \\
          \left(
                 \begin{array}{cc}
           1&  \l(\overline{\Omega(\bar{\lambda})}-\dfrac{\lambda B_1}{\lambda^2+1}\r)e^{-t\Phi} \v\\
               0& 1
                 \end{array}
                 \right), & \lambda \in \overline{D_3}\cap \overline{D_4}, \vspace{0.1in}  \\
          \left(
                 \begin{array}{cc}
            1& \l(\dfrac{b(\lambda)}{\overline{a(\bar{\lambda})}}-\dfrac{\lambda B_1}{\lambda^2+1}\r)e^{-t\Phi} \v\\
               0&1
                 \end{array}
                 \right)
                 \left(
                 \begin{array}{cc}
           1& 0 \v\\
               \l(\dfrac{\lambda \bar{B}_1}{\lambda^2+1}-\dfrac{\overline{b(\bar{\lambda})}}{a(\lambda)}\r)e^{t\Phi}&1
                 \end{array}
                 \right), & \lambda \in \overline{D_4}\cap \overline{D_1},
                 \end{array}
                 \right.$
\end{center}
in terms of the above-mentioned transformation $F^{(1)}(x,t,\lambda)$.

Let
\bee\label{rm}
\begin{array}{l}
h(\lambda)=\d\frac{\lambda \bar{B}_1}{\lambda^2+1}-\Omega(\lambda) , \ \ \ \ \lambda\in \overline{D_2},\v\\
r_1(\lambda)=\d\frac{\overline{b(\bar{\lambda})}}{a(\lambda)}-\frac{\lambda \bar{B}_1}{\lambda^2+1} , \ \ \ \ \lambda\in \overline{D_1}\cap\overline{D_4},\v\\
r(\lambda)=h(\lambda)+r_1(\lambda)= \d\frac{\overline{b(\bar{\lambda})}}{a(\lambda)}-\Omega(\lambda), \ \ \ \ \lambda\in \overline{D_2}\cap\overline{D_3},
\end{array}
\ene
It follows from Assumption 2.1 that we obtain $B_1=b_1$ such that the jump matrix $J^{(1)}(x,t,\lambda)$ has the property that the off-diagonal entries are $O(\lambda^{-2})$ as
$\lambda\to \infty$.

\v
\noindent \textbf{Proposition \, 3.1}
\begin{itemize}
\item {} The functions $h(\lambda)$ is smooth and bounded on $\overline{D_2}$ and analytic in $D_2$ with
\bee
h(\lambda)=\sum_{j=2}^{N}\frac{h_j}{\lambda^j}+O\l(\frac{1}{\lambda^{N+1}}\r) ,  \ \ \ \ \lambda\rightarrow\infty,\quad \lambda\in \overline{D_2} ;
\ene
\item{} The functions $r_1(\lambda)$ is smooth and bounded on $\overline{D_1}\cap\overline{D_4}$;

\item{} The functions $r(\lambda)$ is smooth and bounded on $\overline{D_2}\cap\overline{D_3}$ and analytic in $D_2\cap D_3$.


\end{itemize}

Then the above-mentioned jump matrix $J^{(1)}(x,t,\lambda)$ can be simplified as
\begin{center}
$J^{(1)}(x,t,\lambda)=\left\{
                 \begin{array}{ll}
          \left(
                 \begin{array}{cc}
           1& 0 \v\\
               h(\lambda)e^{t\Phi}& 1
                 \end{array}
                 \right), & \lambda \in \overline{D_1}\cap \overline{D_2}, \vspace{0.1in}  \\
           \left(
                 \begin{array}{cc}
           1& 0\v\\
               r(\lambda)e^{t\Phi}&1
                 \end{array}
                 \right)
                 \left(
                 \begin{array}{cc}
           1& -\overline{r(\bar{\lambda})}e^{-t\Phi} \v\\
               0&1
                 \end{array}
                 \right),
                 & \lambda \in \overline{D_2}\cap \overline{D_3}, \vspace{0.1in}  \\
          \left(
                 \begin{array}{cc}
           1&  -\overline{h(\bar{\lambda})}e^{-t\Phi} \v\\
               0& 1
                 \end{array}
                 \right), & \lambda \in \overline{D_3}\cap \overline{D_4}, \vspace{0.1in}  \\
          \left(
                 \begin{array}{cc}
            1& \overline{r_1(\bar{\lambda})}e^{-t\Phi} \v\\
               0&1
                 \end{array}
                 \right)
                 \left(
                 \begin{array}{cc}
           1& 0 \v\\
               -r_1(\lambda)e^{t\Phi}&1
                 \end{array}
                 \right), & \lambda \in \overline{D_4}\cap \overline{D_1}.
                 \end{array}
                 \right.$
\end{center}

\v
\noindent {\it 3.2.\,  The second modification}
\v

The purpose of the second modification is to deform the vertical part of $\Sigma$ so that it passes through the critical points $\{\lambda_0 ,-\lambda_0,i\lambda_0 ,-i\lambda_0\}$ with
\bee
\lambda_0=\sqrt[4]{\frac{1}{4(\frac{x}{t}+1)}},\qquad    \sqrt[4]{\frac{1}{4(N+1)}}<\lambda_0<\frac{\sqrt{2}}{2},
\ene
which is obtained by solving $\frac{\partial \Phi}{\partial \lambda}=0$ (see Figure 3).

\begin{figure}[!t]
\begin{center}
\vspace{0.05in}
{\scalebox{0.36}[0.3]{\includegraphics{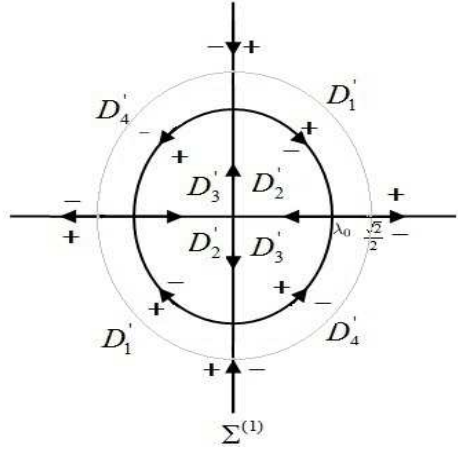}}}
\end{center}
\vspace{-0.2in}\caption{The jump contour $\Sigma^{(1)}$ of the second modification.}
\end{figure}

Now we transform the matrix $M^{(1)}(x, t,\lambda)$ in the first-modification RH problem (\ref{orh1}) by introducing the sectionally analytic function $M^{(2)}(x, t,\lambda)$ by
\bee
M^{(1)}(x, t,\lambda)=M^{(2)}(x, t,\lambda)F^{(2)}(x, t,\lambda),
\ene
where the transformation is defined by
\begin{center}
$F^{(2)}(x,t,\lambda)=\left\{
                 \begin{array}{ll}
          \left(
                 \begin{array}{cc}
           1& 0 \v\\
               -h(\lambda)e^{t\Phi}& 1
                 \end{array}
                 \right), & \lambda \in D'_1 \setminus D_1, \vspace{0.1in}  \\
        \left(
                 \begin{array}{cc}
           1& -\overline{h(\bar{\lambda})}e^{-t\Phi}\v\\
               0& 1
                 \end{array}
                 \right),  & \lambda \in D'_4 \setminus D_4, \vspace{0.1in}  \\
          I& {\rm elsewhere}.
                 \end{array}
                 \right.$
\end{center}
Since $h(\lambda)e^{t\Phi}$ and $\overline{h(\bar{\lambda})}e^{-t\Phi}$ are both bounded and analytic functions of $\lambda \in D'_1\setminus D_1$ and
 $\lambda \in D'_4\setminus D_4$, respectively, thus we know that $M^{(1)}(x,t,\lambda)$ satisfies the  RH problem (\ref{orh1}) if and only if $M^{(2)}(x,t,\lambda)$ solves the second-modification RH problem
\bee\label{orh3}
\left\{\begin{array}{l}
 M^{(2)}(x,t,\lambda) {\rm \,\,is \,\, in\,\, general\,\, a\,\, meromorphic \,\,function\,\, in\,\,} \lambda\in\mathbb{C}\backslash\Sigma^{(1)}, \v\\
 M_{+}^{(2)}(x,t,\lambda)=M_{-}^{(2)}(x,t,\lambda)J^{(2)} (x,t,\lambda) \,\, {\rm \,\,for\,\,} \lambda\in \overline{D'_i}\cap \overline{D'_j},\,\, \,\, i,j=1,2,3,4, \v\\
M^{(2)}(x,t,\lambda)=I+O(\frac{1}{\lambda}) {\rm\,\, as\,\,} \lambda\rightarrow\infty.
\end{array}
\right.
\ene
where the jump matrix $J^{(2)}=F^{(2)}_-J^{(1)} (F^{(2)})^{-1}_+$ is defined by
\begin{center}
$J^{(2)}(x,t,\lambda)=\left\{
                 \begin{array}{ll}
          \left(
                 \begin{array}{cc}
           1& 0\v\\
               h(\lambda)e^{t\Phi}& 1
                 \end{array}
                 \right), & \lambda \in \overline{D'_1}\cap \overline{D'_2}, \vspace{0.1in}  \\
           \left(
                 \begin{array}{cc}
           1& 0\v\\
               r(\lambda)e^{t\Phi}&1
                 \end{array}
                 \right)
                 \left(
                 \begin{array}{cc}
           1& -\overline{r(\bar{\lambda})}e^{-t\Phi}\v\\
               0&1
                 \end{array}
                 \right),
                 & \lambda \in \overline{D'_2}\cap \overline{D'_3}, \vspace{0.1in}  \\
          \left(
                 \begin{array}{cc}
           1&  -\overline{h(\bar{\lambda})}e^{-t\Phi}\v\\
               0& 1
                 \end{array}
                 \right), & \lambda \in \overline{D'_3}\cap \overline{D'_4}, \vspace{0.1in}  \\
          \left(
                 \begin{array}{cc}
            1& \overline{r_1(\bar{\lambda})}e^{-t\Phi}\v\\
               0&1
                 \end{array}
                 \right)
                 \left(
                 \begin{array}{cc}
           1& 0\v\\
               -r_1(\lambda)e^{t\Phi}&1
                 \end{array}
                 \right), & \lambda \in \overline{D'_4}\cap \overline{D'_1},
                 \end{array}
                 \right.$
\end{center}

\v
\noindent {\it 3.3. \, The third modification} \v

\begin{figure}[!t]
\begin{center}
\vspace{0.05in}
{\scalebox{0.38}[0.3]{\includegraphics{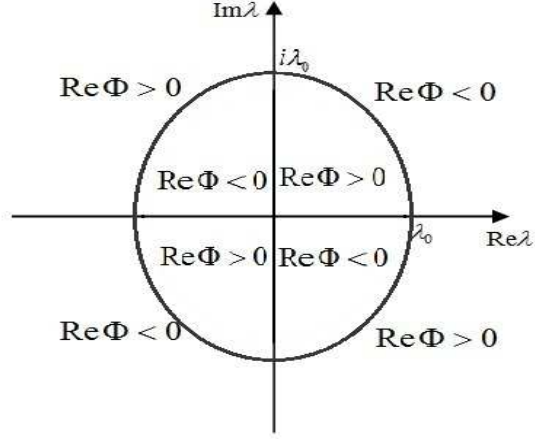}}}
\end{center}
\vspace{-0.25in}\caption{The signature table of $\Re\Phi$.}
\end{figure}

We find that the jump matrix $J^{(2)}(x,t,\lambda)$ has the wrong factorization for $\lambda \in \overline{D'_2}\cap \overline{D'_3}$. Therefore we introduce $M^{(3)}(x,t,\lambda)$ by
\bee
M^{(2)}(x,t,\lambda)=M^{(3)}(x,t,\lambda)F^{(3)}(x,t,\lambda),
\ene
where $F^{(3)}(x,t,\lambda)=\delta^{\sigma_3}(\lambda)$ with $\delta(\lambda)$ satisfying the scalar RH problem
\bee\left\{\begin{array}{ll}
   \delta_{+}(\lambda)=\d\delta_{-}(\lambda)\frac{1}{1-r(\lambda)\overline{r(\bar{\lambda})}},& \lambda \in \overline{D'_2}\cap \overline{D'_3}, \vspace{0.1in} \\
      \delta_{+}(\lambda)=\delta_{-}(\lambda),  & \lambda \in \mathbb{C} \setminus \overline{D'_2}\cap \overline{D'_3}, \vspace{0.1in} \\
      \delta(\lambda)\rightarrow 1,& \lambda\rightarrow\infty,
                 \end{array}\right.
                 \ene
whose solution can be expressed by the formula
\bee\label{delta} \begin{array}{rl}
\delta(\lambda)=&\d\exp\left\{\dfrac{1}{2\pi i}\int_{\overline{D'_2}\cap \overline{D'_3}} \dfrac{-\ln(1-r(\lambda')\overline{r(\bar{\lambda'})})}{\lambda'-\lambda}d\lambda'\right\},\v\\
=&\d\left(\frac{\lambda-\lambda_0}{\lambda}\frac{\lambda+\lambda_0}{\lambda}\right)^{-i\upsilon}e^{-\chi_+(\lambda)-\chi_-(\lambda)}
\left(\frac{\lambda}{\lambda-i\lambda_0}\frac{\lambda}{\lambda+i\lambda_0}\right)^{-i\tilde{\upsilon}}
e^{-\tilde{\chi}_+(\lambda)-\tilde{\chi}_-(\lambda)},
\end{array}
\ene
where
\bee\begin{array}{l}
\upsilon=-\d\frac{1}{2\pi}\ln(1-|r(\lambda_0)|^2), \v\\
\tilde{\upsilon}=-\d\frac{1}{2\pi}\ln(1+|r(i\lambda_0)|^2),\v\\
\chi_{\pm}(\lambda)=\d\frac{1}{2\pi i}\int_{0}^{\pm\lambda_0} \ln \left(\frac{1-|r(\lambda')|^2}{1-|r(\lambda_0)|^2} \right) \frac{\mathrm{d}\lambda'}{\lambda'-\lambda},\v\\
\tilde{\chi}_{\pm}(\lambda)=\d\frac{1}{2\pi i}\int_{\pm i\lambda_0}^{i0} \ln \left(\frac{1-r(\lambda')\overline{r(\overline{\lambda'})}}{1+|r(i\lambda_0)|^2} \right) \frac{\mathrm{d}\lambda'}{\lambda'-\lambda}
\end{array}
\ene
for all $\lambda\in \mathbb{C}$, $|\delta|$ and $|\delta^{-1}|$ are bounded (see Ref.~\cite{xu15}).

Thus we know that $M^{(2)}(x,t,\lambda)$ satisfies the second-modification RH problem (\ref{orh3}) if and only if $M^{(3)}(x,t,\lambda)$ solves the third-modification RH problem
\bee\label{orh4}
\left\{\begin{array}{l}
 M^{(3)}(x,t,\lambda) {\rm \,\,is \,\, in\,\, general\,\, a\,\, meromorphic \,\,function\,\, in\,\,} \lambda\in\mathbb{C}\backslash\Sigma^{(1)}, \v\\
 M_{+}^{(3)}(x,t,\lambda)=M_{-}^{(3)}(x,t,\lambda)J^{(3)} (x,t,\lambda) {\rm \,\,for\,\,} \lambda\in \overline{D'_i}\cap \overline{D'_j},\,\, i,j=1,2,3,4, \v\\
M^{(3)}(x,t,\lambda)=I+O(\frac{1}{\lambda}) {\rm\,\, as\,\,} \lambda\rightarrow\infty.
\end{array}
\right.
\ene
where the jump matrix  $J^{(3)}=F^{(3)}_-J^{(2)} (F^{(3)})^{-1}_+$ defined by
\begin{center}
$J^{(3)}(x,t,\lambda)=\left\{
                 \begin{array}{ll}
          \left(
                 \begin{array}{cc}
           1& 0\v\\
               h(\lambda)\delta^{-2}e^{t\Phi}& 1
                 \end{array}
                 \right), & \lambda \in \overline{D'_1}\cap \overline{D'_2}, \vspace{0.1in}  \\
           \left(
                 \begin{array}{cc}
           1& -\overline{r_2(\bar{\lambda})}\delta_-^{2}e^{-t\Phi}\v\\
               0&1                \end{array}
                 \right)
           \left(
                 \begin{array}{cc}
           1& 0\v\\
               r_2(\lambda)\delta_+^{-2}e^{t\Phi}&1
                 \end{array}
                 \right),
                 & \lambda \in \overline{D'_2}\cap \overline{D'_3}, \vspace{0.1in}  \\
          \left(
                 \begin{array}{cc}
           1&  -\overline{h(\bar{\lambda})}\delta^{2}e^{-t\Phi}\v\\
               0& 1
                 \end{array}
                 \right), & \lambda \in \overline{D'_3}\cap \overline{D'_4}, \vspace{0.1in}  \\
          \left(
                 \begin{array}{cc}
            1& \overline{r_1(\bar{\lambda})}\delta^{2}e^{-t\Phi}\v\\
               0&1
                 \end{array}
                 \right)
                 \left(
                 \begin{array}{cc}
           1& 0 \v\\
               -r_1(\lambda)\delta^{-2}e^{t\Phi}&1
                 \end{array}
                 \right), & \lambda \in \overline{D'_4}\cap \overline{D'_1},
                 \end{array}
                 \right.$
\end{center}
where we have introduced $r_2(\lambda)$ in the form
\bee
 r_2(\lambda)=\frac{r(\lambda)}{1-r(\lambda)\overline{r(\bar{\lambda})}}.
\ene

\begin{figure}[!t]
\begin{center}
\vspace{0.05in}
{\scalebox{0.438}[0.33]{\includegraphics{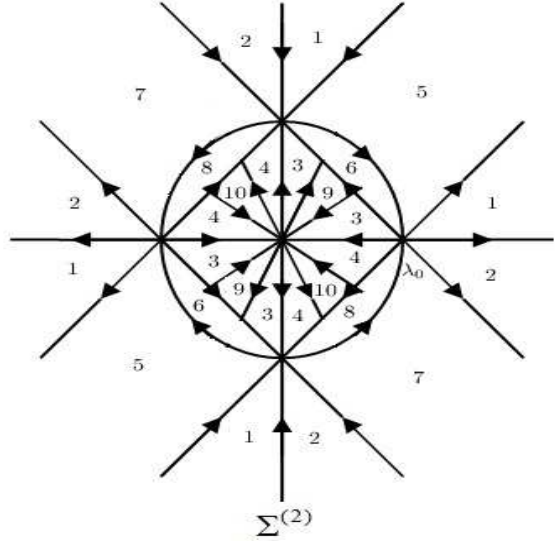}}}
\end{center}
\vspace{-0.25in}\caption{The jump contour $\Sigma^{(2)}$ of the forth modification.}
\end{figure}

\v
\noindent {\it 3.4. \,  The forth modification} \v

The aim of the forth modification is to distort the contour $\Sigma^{(2)}$ (see Figure 5) such that the jump matrix contains
the exponential factor $e^{-t\Phi}$ on the parts of the contour where ${\rm Re}\,\Phi >0$ , and the factor $e^{t\Phi}$ on the parts where ${\rm Re}\,\Phi<0$. Decompose each of the functions $h(\lambda)$,\,$r_1(\lambda)$, \,$r_2(\lambda)$ into an analytic part and a small remainder, respectively. As a consequence, this transformation can distort the analytic parts of the jump matrix, whereas the small remainder will be left on the previous contour.

\v
\noindent \textbf{Proposition \, 3.2}\,
There exist the following decompositions
\bee\begin{array}{l}
h(\lambda)=h_a(t,\lambda)+h_r(t,\lambda),  \quad t>0, \quad \lambda\in \overline{D'_2}\cap\overline{D'_1}, \v\\
r_1(\lambda)=r_{1a}(t,\lambda)+r_{1r}(t,\lambda),  \quad t>0, \quad \lambda\in (-\infty,-\lambda_0)\cup (\lambda_0,\infty),\v\\
r_2(\lambda)=r_{2a}(t,\lambda)+r_{2r}(t,\lambda), \quad t>0,  \quad  \lambda\in (-\lambda_0,\lambda_0)\cup i(-\lambda_0,\lambda_0),
\end{array}
\ene
where the functions $h_a(t,\lambda)$,\, $h_r(t,\lambda)$,\, $r_{ja}(t,\lambda)$, and $r_{jr}(t,\lambda)\, (j=1,2)$ have the following properties
\begin{itemize}
\item{} For each $t>0$, $h_a(t,\lambda)$ is defined and continuous for $\lambda\in \overline{D'_1}$ and analytic for $\lambda\in D_1'$ ;
\item{} The functions $h_a(t,\lambda)$ satisfies
\bee
|h_a(t,\lambda)|\leq \frac{c}{1+|\lambda|^2}e^{\frac{t}{4}|{\rm Re}\Phi(\varsigma,\lambda)|} ,  \quad  \lambda\in D_1',\quad 0<\frac{x}{t}\triangleq\varsigma<N;
\ene
\item{} The $L^1$, $L^2$, and $L^\infty$ norms of the function $h_r(t,\lambda)$ on $\overline{D_1'}\cap\overline{D_2'}$ are $O(t^{-\frac{3}{2}})$ as $t\rightarrow\infty$;

\item{} For each $t>0$, $r_{1a}(t,\lambda)$ is defined and continuous for $\lambda\in \overline{D'_1}$ and analytic for $\lambda\in D_1'$;

\item{} The functions $r_{1a}(t,\lambda)$ satisfies
\bee
|r_{1a}(t,\lambda)|\leq \frac{c}{1+|\lambda|^2}e^{\frac{t}{4}|{\rm Re}\Phi(\varsigma,\lambda)|} , \quad
  \lambda\in D_1', \quad 0<\frac{x}{t}\triangleq\varsigma<N;
\ene
\item{} The $L^1$, $L^2$, and $L^\infty$ norms of the function $r_{1r}(t,\lambda)$ on $\lambda\in (-\infty,-\lambda_0)\cup (\lambda_0,\infty)$ are $O(t^{-\frac{3}{2}})$ as $t\rightarrow\infty$;

\item{} For each $t>0$, $r_{2a}(t,\lambda)$ is defined and continuous for $\lambda\in \overline{D'_3}$ and analytic for $\lambda\in D_3'$;

\item{} The functions $r_{2a}(t,\lambda)$ satisfies
\bee
|r_{2a}(t,\lambda)|\leq \frac{c}{1+|\lambda|^2}e^{\frac{t}{4}|{\rm Re}\Phi(\varsigma,\lambda)|} , \quad  \lambda\in D_3',\quad 0<\frac{x}{t}\triangleq\varsigma<N;
\ene

\item{} The $L^1$, $L^2$, and $L^\infty$ norms of the function $r_{2r}(t,\lambda)$ on $\lambda\in (-\lambda_0,\lambda_0)\cup i(-\lambda_0,\lambda_0)$ are $O(t^{-\frac{3}{2}})$ as $t\rightarrow\infty$.
\end{itemize}

\noindent {\bf Proof.} We only show the propositions of $h(\lambda)$ here, and the proofs of $r_{1}(\lambda)$.
 Similarly to Ref.~\cite{ibv8}, since
$h(\lambda)\in \mathbb{C}^\infty(\overline{D'_2}\cap\overline{D'_1})$, then we have
\bee\begin{array}{l}
h^{(n)}(\lambda)=\d\frac{\mathrm{d}^n}{\mathrm{d}\lambda^n}\left(\sum_{j=0}^{4}p_j\lambda^j\right)+O(\lambda^{5-n}), \quad
\lambda\rightarrow 0, \quad \lambda \in \overline{D'_2}\cap\overline{D'_1}, \quad n=0,1,2, \v\\
h^{(n)}(\lambda)=\d\frac{\mathrm{d}^n}{\mathrm{d}\lambda^n}\left(\sum_{j=2}^{3}\frac{h_j}{\lambda^{j}}\right)+O(\lambda^{-4-n}), \quad
\lambda\rightarrow\infty, \quad \lambda \in \overline{D'_2}\cap\overline{D'_1}, \quad n=0,1,2.
\end{array}
\ene

Let
\bee
f_0(\lambda)=\sum_{j=2}^{8}\frac{a_j}{(\lambda+i)^{j}},
\ene
where $\{a_j\}_{2}^{8}$ are complex constants satisfy
\bee \label{f0}
f_0(\lambda)=\left\{
                 \begin{array}{cc}
         \d \sum_{j=0}^{4}p_j\lambda^j+O(\lambda^5),& \lambda \rightarrow 0, \vspace{0.1in}  \\
          \d  \sum_{j=2}^{3}h_j\lambda^{-j}+O(\lambda^{-4}), & \lambda \rightarrow\infty.
                 \end{array}
                 \right.
\ene
It is easy to verify that Eq.~(\ref{f0}) imposes seven linearly independent conditions on constants $a_j$, hence the coefficients $a_j$  exist and unique.

Let
\bee
f(\lambda)=h(\lambda)-f_0(\lambda).
\ene
Then we have
\bee
f^{(n)}(\lambda)=\left\{
                 \begin{array}{cc}
          O(\lambda^{5-n}),& \lambda \rightarrow0, \vspace{0.1in}  \\
        O(\lambda^{-4-n}),& \lambda \rightarrow\infty.
                 \end{array}
                 \right.
\ene

Define $\Xi(\lambda)=\omega$ is a bijection $\{\lambda|\lambda=\lambda_0e^{i\omega },\,\, 0<\omega<\frac{\pi}{2} \ \ {\rm or}  \ \ \pi<\omega<\frac{3\pi}{2}\}\rightarrow \mathbb{R}$ , and $|\Re (i\Xi)|<|\Re(\Phi)|$ for $\lambda\in D_1$.
Let $G(\Xi)=(\lambda+i)^2f(\lambda)$, then we have
\bee
G^{(j)}(\Xi)=\left(\frac{1}{\Xi'(\lambda)}\frac{\partial}{\partial\lambda}\right)^j(\lambda+i)^2f(\lambda).
\ene
Since $||G^{(j)}(\Xi)||_{L^2(\mathbb{R})}<\infty$,\, $j=0,1,2$, thus $G^{(j)}(\Xi)$ belongs to the Sobolev space $H^2(\mathbb{R})$, which leads to
\bee
||s^2\hat{G}(s)||_{L^2(\mathbb{R})}<\infty
\ene
with $\hat{G}(s) =\frac{1}{2\pi}\int_\mathbb{R} G(\Xi) e^{-i\Xi s}\mathrm{d}\Xi$.

As a result, we have $G(\Xi) =\int_\mathbb{R} \hat{G}(s) e^{i\Xi s}\mathrm{d}s=(\lambda+i)^2f(\lambda)$, that is
\bee\begin{array}{rl}
f(\lambda)=&\d \frac{1}{(\lambda+i)^2}\int_\mathbb{R} \hat{G}(s) e^{i\Xi s}\mathrm{d}s \v\\
=&\d \frac{1}{(\lambda+i)^2}\int_{-\infty}^{-\frac{t}{4}} \hat{G}(s) e^{i\Xi s}\mathrm{d}s+\frac{1}{(\lambda+i)^2}\int_{-\frac{t}{4}}^{\infty} \hat{G}(s) e^{i\Xi s}\mathrm{d}s \v\\
=&f_r(t,\lambda)+f_a(t,\lambda),
\end{array}
\ene

We further know that
\bee\begin{array}{rl}
f_r(t,\lambda)\leq &\d\frac{1}{|\lambda+i|^2}\int_{-\infty}^{-\frac{t}{4}} s^2|\hat{G}(s)| s^{-2}\mathrm{d}s\leq \frac{c}{1+|\lambda|^2}t^{-\frac{3}{2}},\,\, t>0,\,\, \lambda\in \overline{D_1'}\cap \overline{D_2'}, \v\\
f_a(t,\lambda)\leq& \d\frac{1}{|\lambda+i|^2}||\hat{G}(s)||_{L^1(\mathbb{R})} \sup_{s>\frac{-t}{4}}e^{s\,\Re(i\Xi)}\leq \frac{c}{1+|\lambda|^2}e^{\frac{t}{4}|\Re(i\Xi)|} \v\\
 \leq & \d\frac{c}{1+|\lambda|^2}e^{\frac{t}{4}|\Re(\Phi)|},\,\, t>0,\,\, \lambda\in D_1'.
\end{array}
\ene

Therefore we have
\bee\begin{array}{rl}
h_a(t,\lambda)=&\!\!\! f_0(\lambda)+f_a(t,\lambda)\leq\d \frac{c}{1+|\lambda|^2}e^{\frac{t}{4}|{\rm Re}\Phi(\zeta,\lambda)|},\,\, t>0,\,\,\lambda\in D_1', \v\\
h_r(t,\lambda)=&\!\!\! f_r(t,\lambda)=O(t^{-\frac{3}{2}}),\,\, t>0,\,\, \lambda\in \overline{D_1'}\cap \overline{D_2'}.
\end{array}
\ene
This completes the proof of the properties of $h(\lambda)$. $\Box$ \\

Therefore, we introduce $M^{(4)}(x,t,\lambda)$ by
\bee
M^{(3)}(x,t,\lambda)=M^{(4)}(x,t,\lambda)F^{(4)}(x,t,\lambda),
\ene
where the transformation is given as
\begin{center}
$F^{(4)}(x,t,\lambda)=\left\{
                 \begin{array}{ll}
          \left(
                 \begin{array}{cc}
           1& 0\v\\
               -r_{1a}(\lambda)\delta^{-2}e^{t\Phi}& 1
                 \end{array}
                 \right), & \lambda \in 1, \vspace{0.1in}  \\
           \left(
                 \begin{array}{cc}
           1& -\overline{r_{1a}(\bar{\lambda})}\delta^{2}e^{-t\Phi}\v\\
               0& 1
                 \end{array}
                 \right), & \lambda \in 2, \vspace{0.1in}  \\
         \left(
                 \begin{array}{cc}
           1& \overline{r_{2a}(\bar{\lambda})}\delta^{2}e^{-t\Phi}\v\\
               0& 1
                 \end{array}
                 \right), & \lambda \in 3, \vspace{0.1in}  \\
         \left(
                 \begin{array}{cc}
           1& 0\v\\
               r_{2a}(\lambda)\delta^{-2}e^{t\Phi}& 1
                 \end{array}
                 \right), & \lambda \in 4, \vspace{0.1in}  \\
        \left(
                 \begin{array}{cc}
           1& 0\v\\
               h_{a}(\lambda)\delta^{-2}e^{t\Phi}& 1
                 \end{array}
                 \right), & \lambda \in 5, \vspace{0.1in}  \\
          \left(
                 \begin{array}{cc}
           1& \overline{h_{a}(\bar{\lambda})}\delta^{2}e^{-t\Phi}\v\\
               0& 1
                 \end{array}
                 \right), & \lambda \in 7, \vspace{0.1in}  \\
          I& \lambda\in 6,8,9,10,
                 \end{array}
                 \right.$
\end{center}

 We know that $M^{(3)}(x,t,\lambda)$ satisfies the third-modification  RH problem (\ref{orh4}) if and only if $M^{(4)}(x,t,\lambda)$ solves the
 forth-modification RH problem
\bee
\left\{\begin{array}{l}
 M^{(4)}(x,t,\lambda) {\rm \,\,is \,\, in\,\, general\,\, a\,\, meromorphic \,\,function\,\, in\,\,} \lambda\in\mathbb{C}\backslash\Sigma^{(2)}, \v\\
 M_{+}^{(4)}(x,t,\lambda)=M_{-}^{(4)}(x,t,\lambda)J^{(4)} (x,t,\lambda) \,\, {\rm \,\,for\,\,} \lambda\in i\cap j,\quad i,j=1,2,...,10, \v\\
M^{(4)}(x,t,\lambda)=I+O(\frac{1}{\lambda}) {\rm\,\, as\,\,} \lambda\rightarrow\infty.
\end{array}
\right.
\ene
where the jump matrix $J^{(4)}(x,t,\lambda)=F^{(4)}_-(x,t,\lambda)J^{(3)}(x,t,\lambda) (F^{(4)})^{-1}_+(x,t,\lambda)$ is defined by
\bee
J^{(4)}(x,t,\lambda)=\left\{\begin{array}{ll}
          \left(\begin{array}{cc} 1& 0 \v\\  h_r(\lambda)\delta^{-2}e^{t\Phi}& 1 \end{array}
               \right), & \lambda \in 5\cap6, \vspace{0.1in}  \\
          \left(\begin{array}{cc}
           1& -\overline{r_{2r}(\bar{\lambda})}\delta_-^2e^{-t\Phi} \v\\  0&1 \end{array} \right)
           \left(\begin{array}{cc}  1& 0\\   r_{2r}(\lambda)\delta_+^{-2}e^{t\Phi}&1 \\ \end{array}\right),
            & \lambda \in 3\cap4, \vspace{0.1in}  \\
          \left( \begin{array}{cc}
           1&  -\overline{h_r(\bar{\lambda})}\delta^2e^{-t\Phi} \v\\        0& 1        \end{array} \right),
             & \lambda \in 7\cap8, \vspace{0.1in}  \\
          \left( \begin{array}{cc}
            1& \overline{r_{1r}(\bar{\lambda})}\delta^2e^{-t\Phi} \v\\          0&1   \end{array}      \right)
                 \left(  \begin{array}{cc}   1& 0 \v\\       -r_{1r}(\lambda)\delta^{-2}e^{t\Phi}&1  \end{array} \right),
                  & \lambda \in 1\cap2, \vspace{0.1in}  \\
           \left(  \begin{array}{cc}
                  1& -\overline{r_{2a}(\bar{\lambda})}\delta^2e^{-t\Phi} \v\\              0&1   \end{array}     \right),
                   & \lambda \in 3\cap6, \vspace{0.1in}  \\
            \left(                 \begin{array}{cc}          1& 0\v\\      r_{2a}(\lambda)\delta^{-2}e^{t\Phi}&1
                 \end{array}     \right), & \lambda \in 4\cap8, \vspace{0.1in}  \\
          \left(
                 \begin{array}{cc}
           1& 0 \v\\
              -(r_{1a}(\lambda)+h_a(\lambda))\delta^{-2}e^{t\Phi}&1
                 \end{array}
                 \right) & \lambda \in 1\cap5, \vspace{0.1in}  \\
        \left(
                 \begin{array}{cc}
           1& (\overline{r_{1a}(\bar{\lambda})}+\overline{h_a(\bar{\lambda})})\delta^{2}e^{-t\Phi} \v\\
             0 &1     \end{array}  \right), & \lambda \in 2\cap7, \vspace{0.1in}  \\
        \left(\begin{array}{cc}
           1& -\overline{r_{2a}(\bar{\lambda})}\delta^2e^{-t\Phi} \v\\
             0 &1                  \end{array}
                 \right), & \lambda \in 3\cap9, \vspace{0.1in}  \\
                   \left(\begin{array}{cc}
           1& 0 \v\\
           r_{2a}(\lambda)\delta^{-2}e^{t\Phi}  &1   \end{array}
                 \right), & \lambda \in 4\cap10,
                \end{array}     \right.
\ene

{\bf Remark.} The above-mentioned transformations change a RH problem for $M^{(3)}(x,t,\lambda)$ with the property that
the jump matrix $J^{(4)}(x,t,\lambda)$ decays to $I$ as $t\to\infty$ everywhere except near the critical points
$\{\lambda_0 ,-\lambda_0,i\lambda_0 ,-i\lambda_0\}$. This implies that we only need to consider
a neighborhood of the critical points $\{\lambda_0 ,-\lambda_0,i\lambda_0 ,-i\lambda_0\}$ when we studying
the long-time asymptotics of $M^{(4)}(x,t,\lambda)$ in terms of the corresponding RH problem.

\begin{figure}[!t]
\begin{center}
\vspace{0.05in}
{\scalebox{0.3}[0.25]{\includegraphics{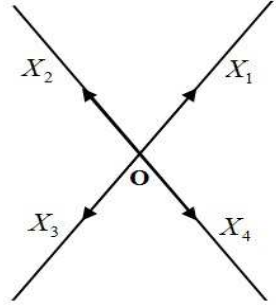}}}
\end{center}
\vspace{-0.6in}\caption{The contour $X=X_{1}\cup X_{2}\cup X_{3} \cup X_{4}$.}
\label{x-region}
\end{figure}

\section{The local model nearby critical points}

\v
\noindent {\it 4.1.\,  Modelling the RH problem}

Let $X$ denote the cross defined by $X=X_{1}\cup X_{2}\cup X_{3} \cup X_{4}\subset \mathbb{C}$ with $X_j$ given by (see Figure 6)
\bee\begin{array}{l}
X_{1}=\{le^{\frac{\pi}{4}i}|\,0\leq l<\infty\}, \ \ \ \ X_{2}=\{le^{\frac{3\pi}{4}i}|\,0\leq l<\infty\}, \v\\
X_{3}=\{le^{\frac{-3\pi}{4}i}|\,0\leq l<\infty\}, \ \ \ X_{4}=\{le^{\frac{-\pi}{4}i}|\,0\leq l<\infty\}.
\end{array}
\ene

Let $\mathbb{D}\subset\mathbb{C}$ denote the open unit disk and define the functions $\upsilon(p)({\rm or} \,\, \tilde{\upsilon}(p)):\mathbb{D}\rightarrow(0,\infty)$ by
$\upsilon(p)=-\frac{1}{2\pi}{\rm ln}(1-|p|^2)({\rm or}\,\, \tilde{\upsilon}(p)=-\frac{1}{2\pi}{\rm ln}(1+|p|^2))$.
Consider the following RH problem parameterized by $p\in\mathbb{D}$. Following the properties in Refs.~\cite{le15,its}, we have the following Lemma:

\noindent \textbf{Lemma 4.1}\,
{\it Case 1.} Consider the following RH problem
\begin{eqnarray}\label{zhenggui}
\left\{
                 \begin{array}{ll}
              M^X_{+}(p,z)=M^X_{-}(p,z)J^X(p,z), &  {\rm for} \,\, a.e. \ \  z\in X, \vspace{0.1in} \\
              M^X(p,z) \rightarrow I, &z\rightarrow \infty,
                 \end{array}
                 \right.
\end{eqnarray}
where the jump matrix $J^X(p,z)$ is defined by
\bee
J^X(p,z)=\left\{
                 \begin{array}{ll}
                 \left(\begin{array}{cc}        1& 0 \vspace{0.05in}\\
               -p(\varsigma)z^{2i\upsilon(p)}e^{\frac{iz^2}{2}}& 1
                 \end{array}
                 \right), & z\in X_{1}, \vspace{0.1in} \\
          \left(
                 \begin{array}{cc}
           1& \dfrac{-\bar{p}(\varsigma)}{1- |p(\varsigma)|^2}z^{-2i\upsilon(p)}e^{-\frac{iz^2}{2}} \vspace{0.05in}\\
                0& 1
                 \end{array}
                 \right), & z\in X_{2}, \vspace{0.1in}  \\
                \left(
                 \begin{array}{cc}
                         1& 0 \vspace{0.05in}\\
                \dfrac{ p(\varsigma)}{1-|p(\varsigma)|^2}z^{2i\upsilon(p)}e^{\frac{iz^2}{2}}& 1
                 \end{array}
                 \right), & z\in X_{3},\vspace{0.1in} \\
                  \left(
                 \begin{array}{cc}
           1& \bar{p}(\varsigma) z^{-2i\upsilon(p)}e^{-\frac{iz^2}{2}} \vspace{0.05in}\\
                0& 1
                 \end{array}
                 \right), &z\in X_{4}.
                 \end{array}
                 \right.
                 \ene
The matrix $J^X(p,z)$ has entries that oscillate rapidly as $z\rightarrow 0$ and $J^X(p,z)$ is not continuous at $z=0$, but $J^X(p,z)-I\in L^2(X)\cap L^\infty(X)$. Thus the RH problem of Eq.~(\ref{zhenggui}) has a unique solution and can be solved explicitly in terms of parabolic cylinder functions as
\bee
M^X(p,z)=I-\frac{i}{z}\left(
                 \begin{array}{cc}
           0& \beta^X(p) \vspace{0.05in}\\
              \overline{\beta^X(p)} & 0
                 \end{array}
                 \right)+O\left(\frac{p}{z^2}\right), \, \, z\rightarrow\infty.
\ene
where
\begin{center}
$\beta^{X}(p)=\sqrt{\upsilon(p)}e^{i[\frac{\pi}{4}+\arg p-\arg \Gamma(-i\upsilon(p))]}$
\end{center}

{\it Case 2.} Consider the following RH problem
\begin{eqnarray}
\left\{
                 \begin{array}{ll}
              M^Y_{+}(p,z)=M^Y_{-}(p,z)J^Y(p,z), &  {\rm for} \,\, a.e. \ \  z\in X, \vspace{0.1in} \\
              M^Y(p,z) \rightarrow I, &z\rightarrow \infty.
                 \end{array}
                 \right.
\end{eqnarray}
where the jump matrix $J^Y(p,z)$ is defined by
\bee
J^Y(p,z)=\left\{
                 \begin{array}{ll}
                 \left(\begin{array}{cc}        1& 0 \vspace{0.05in}\\
               -p(\varsigma)z^{-2i\tilde{\upsilon}(p)}e^{\frac{iz^2}{2}}& 1
                 \end{array}
                 \right), & z\in X_{1}, \vspace{0.1in} \\
          \left(
                 \begin{array}{cc}
           1& \dfrac{-\bar{p}(\varsigma)}{1- p(\varsigma)\overline{p(\bar{\varsigma})}}z^{2i\tilde{\upsilon}(p)}e^{-\frac{iz^2}{2}} \vspace{0.05in}\\
                0& 1
                 \end{array}
                 \right), & z\in X_{2}, \vspace{0.1in}  \\
                \left(
                 \begin{array}{cc}
                         1& 0 \vspace{0.05in}\\
                \dfrac{ p(\varsigma)}{1-p(\varsigma)\overline{p(\bar{\varsigma})}}z^{-2i\tilde{\upsilon}(p)}e^{\frac{iz^2}{2}}& 1
                 \end{array}
                 \right), & z\in X_{3},\vspace{0.1in} \\
                  \left(
                 \begin{array}{cc}
           1& \overline{p(\bar{\varsigma})} z^{2i\tilde{\upsilon}(p)}e^{-\frac{iz^2}{2}} \vspace{0.05in}\\
                0& 1
                 \end{array}
                 \right), &z\in X_{4}.
                 \end{array}
                 \right.
                 \ene
The matrix $J^Y(p,z)$ has entries that oscillate rapidly as $z\rightarrow 0$ and $J^Y(p,z)$ is not continuous at $z=0$, but $J^Y(p,z)-I\in L^2(X)\cap L^\infty(X)$. Thus the RH problem of Eq.~(\ref{zhenggui}) has a unique solution and can be solved explicitly in terms of parabolic cylinder functions as
\bee
M^Y(p,z)=I-\frac{i}{z}\left(
                 \begin{array}{cc}
           0& \beta^Y(p(\varsigma)) \v\\
              -\overline{\beta^Y(p(\bar{\varsigma}))} & 0
                 \end{array}
                 \right)+O\left(\frac{p}{z^2}\right), \, \, z\rightarrow\infty.
\ene
where
\begin{center}
$\beta^{Y}(p)=\sqrt{\tilde{\upsilon}(p)}e^{i[\frac{\pi}{4}+\arg p(\varsigma)+\arg \Gamma(i\tilde{\upsilon}(p))]}$
\end{center}

\v
\noindent {\it 4.2.\,  Local model nearby critical points}\v

 For a small $\varepsilon >0$, let  $D_\varepsilon(j)$ stand for the open disk of radius $\varepsilon$ centered at the critical points $j=\pm\lambda_0,\pm i\lambda_0$.
  To relate $M^{(4)}$ to the solution $M^X(\varsigma,z)$ of Lemma 4.1, we employ a local change of variables of $\lambda$ near $\pm\lambda_0,\pm i\lambda_0$ and introduce the scaling transformations by (see ~\cite{xu15})
\bee\l\{\begin{array}{rl}
S_{\lambda_0}:&   \lambda \mapsto \d\frac{\lambda^2_0}{2\sqrt t}z+\lambda_0,\v\\
S_{-\lambda_0}:&   \lambda \mapsto \d\frac{\lambda^2_0}{2\sqrt t}z-\lambda_0,\v\\
S_{i\lambda_0}: &  \lambda \mapsto \d\frac{-\lambda^2_0}{2\sqrt t}z+i\lambda_0,\v\\
S_{-i\lambda_0}: &  \lambda \mapsto \d\frac{-\lambda^2_0}{2\sqrt t}z-i\lambda_0.
\end{array}\r.\ene

As a consequence, we have the following properties:

{\it Case 1.} For $S_{\lambda_0}$, we have
\bee
S_{\lambda_0}\delta e^{-\frac{t\Phi}{2}}=\delta^0_{\lambda_0}\delta^1_{\lambda_0},
\ene
where
\bee\no
\begin{array}{rl}
\delta^0_{\lambda_0}=&\!\!\d\frac{{\lambda_0}^{2i\tilde{\upsilon}-i\upsilon}}{(\sqrt t)^{-i\upsilon}}2^{i\tilde{\upsilon}}e^{it-i\frac{t}{2\lambda^2_0}}e^{-\chi_{\pm}(\lambda_0)}e^{-\widetilde{\chi}'_{\pm}(\lambda_0)},\\
\delta^1_{\lambda_0}=&\!\! \d z^{-i\upsilon}e^{-i\frac{z^2}{4}+i\frac{\lambda^6_0z^3}{\xi^5\sqrt t}}\frac{\lambda^{-2i\tilde{\upsilon}-i\upsilon}_0}{2^{i\tilde{\upsilon}-i\upsilon}}\frac{(\frac{\lambda^2_0}{2\sqrt t}z+2\lambda_0)^{-i\upsilon}}{(\frac{\lambda^2_0}{2\sqrt t}z+\lambda_0)^{-2i\upsilon}} \v\\
&\times\d\left[\l(\frac{\lambda^2_0}{2\sqrt t}z+\lambda_0+i\lambda_0\r)\l(\frac{\lambda^2_0}{2\sqrt t}z+\lambda_0-i\lambda_0\r)\right]^{i\tilde{\upsilon}} \v\\
&\d\times e^{-\chi_{\pm}(\frac{\lambda^2_0}{2\sqrt t}z+\lambda_0)+\chi_{\pm}(\lambda_0)}e^{-\tilde{\chi}'_{\pm}(\frac{\lambda^2_0}{2\sqrt t}z+\lambda_0)+\tilde{\chi}'_{\pm}(\lambda_0)},
\end{array}\ene
with
\bee\label{chi}
\tilde{\chi}'_{\pm}(z)=\exp\left[\frac{i}{2\pi}\int_{\pm \lambda_0}^{0} \ln |z-iz'|d\ln(1+|r(iz')|^2)\right].
\ene

{\it Case 2.} For $S_{-\lambda_0}$, we have
\bee
S_{-\lambda_0}\delta e^{-\frac{t\Phi}{2}}=\delta^0_{-\lambda_0}\delta^1_{-\lambda_0},
\ene
where
\bee \no
\begin{array}{rl}
\delta^0_{-\lambda_0}=&\!\!\!\!\d\frac{{\lambda_0}^{2i\tilde{\upsilon}-i\upsilon}}{(\sqrt t)^{-i\upsilon}}2^{i\tilde{\upsilon}}e^{it-i\frac{t}{2\lambda^2_0}}e^{-\chi_{\pm}(-\lambda_0)}e^{-\widetilde{\chi}'_{\pm}(-\lambda_0)},\v\\
\delta^1_{-\lambda_0}=&\!\!\!\!\d(-z)^{-i\upsilon}e^{-i\frac{z^2}{4}+i\frac{-\lambda^6_0z^3}{\xi^5\sqrt t}}\frac{(-\lambda_0)^{-2i\tilde{\upsilon}-i\upsilon}}{2^{i\tilde{\upsilon}-i\upsilon}}\frac{(\frac{\lambda^2_0}{2\sqrt t}z-2\lambda_0)^{-i\upsilon}}{(\frac{\lambda^2_0}{2\sqrt t}z-\lambda_0)^{-2i\upsilon}} \v\\
&\times\d\l[\l(\frac{\lambda^2_0}{2\sqrt t}z-\lambda_0+i\lambda_0\r)\l(\frac{\lambda^2_0}{2\sqrt t}z-\lambda_0-i\lambda_0\r)\r]^{i\tilde{\upsilon}} \v\\
&\d\times e^{-\chi_{\pm}(\frac{\lambda^2_0}{2\sqrt t}z-\lambda_0)+\chi_{\pm}(-\lambda_0)}e^{-\tilde{\chi}'_{\pm}(\frac{\lambda^2_0}{2\sqrt t}z-\lambda_0)+\tilde{\chi}'_{\pm}(-\lambda_0)},
\end{array}
\ene
with $\tilde{\chi}'_{\pm}(z)$ given by Eq.~(\ref{chi}).

{\it Case 3.} For $S_{i\lambda_0}$, we have
\bee
S_{i\lambda_0}\delta e^{-\frac{t\Phi}{2}}=\delta^0_{i\lambda_0}\delta^1_{i\lambda_0},
\ene
where
\bee\no
\begin{array}{rl}
\delta^0_{i\lambda_0}=&\!\!\!\!\d\frac{{\lambda_0}^{-2i\upsilon+i\tilde{\upsilon}}}{(\sqrt t)^{i\tilde{\upsilon}}}2^{-i\upsilon}e^{it+i\frac{t}{2\lambda^2_0}}e^{-\chi'_{\pm}(i\lambda_0)}e^{-\widetilde{\chi}_{\pm}(i\lambda_0)},\v\\\
\delta^1_{i\lambda_0}=&\!\!\!\!\d(iz)^{i\tilde{\upsilon}}e^{-i\frac{z^2}{4}+i\frac{\lambda^6_0z^3}{\xi^5\sqrt t}}\frac{i^{i\tilde{\upsilon}}\lambda^{2i\upsilon+i\tilde{\upsilon}}_0}{2^{i\tilde{\upsilon}+i\upsilon}}\frac{(\frac{-\lambda^2_0}{2\sqrt t}z+i\lambda_0)^{-2i\tilde{\upsilon}}}{(\frac{-\lambda^2_0}{2\sqrt t}z+2i\lambda_0)^{-2i\tilde{\upsilon}}} \v\\
&\d\times \l[\l(\frac{-\lambda^2_0}{2\sqrt t}z+\lambda_0+i\lambda_0\r)\l(\frac{-\lambda^2_0}{2\sqrt t}z+\lambda_0-i\lambda_0\r)\r]^{-i\upsilon} \v\\
&\d\times e^{-\chi'_{\pm}(\frac{-\lambda^2_0}{2\sqrt t}z+i\lambda_0)+\chi'_{\pm}(i\lambda_0)}e^{-\tilde{\chi}_{\pm}(\frac{-\lambda^2_0}{2\sqrt t}z+i\lambda_0)+\tilde{\chi}_{\pm}(i\lambda_0)},
\end{array}
\ene
with
\bee\label{chii}
\chi'_{\pm}(z)=\exp\l[\frac{i}{2\pi}\int_{0}^{\pm \lambda_0} \ln |z-z'|d\ln(1-|r(z')|^2)\r].
\ene
{\it Case 4.} For $S_{-i\lambda_0}$, we have
\bee
S_{-i\lambda_0}\delta e^{-\frac{t\Phi}{2}}=\delta^0_{-i\lambda_0}\delta^1_{-i\lambda_0},
\ene
where
\bee\no
\begin{array}{rl}
\delta^0_{-i\lambda_0}=&\!\!\!\!\d\frac{{\lambda_0}^{-2i\upsilon+i\tilde{\upsilon}}}{(\sqrt t)^{i\tilde{\upsilon}}}2^{-i\upsilon}e^{it+i\frac{t}{2\lambda^2_0}}e^{-\chi'_{\pm}(-i\lambda_0)}
e^{-\widetilde{\chi}_{\pm}(-i\lambda_0)}, \v\\
\delta^1_{-i\lambda_0}=&\!\!\!\!\d(-iz)^{i\tilde{\upsilon}}e^{-i\frac{z^2}{4}+i\frac{\lambda^6_0z^3}{\xi^5\sqrt t}}\frac{(-i)^{i\tilde{\upsilon}}\lambda^{2i\upsilon+i\tilde{\upsilon}}_0}{2^{i\tilde{\upsilon}+i\upsilon}}\frac{(\frac{-\lambda^2_0}{2\sqrt t}z-i\lambda_0)^{-2i\tilde{\upsilon}}}{(\frac{-\lambda^2_0}{2\sqrt t}z-2i\lambda_0)^{-2i\tilde{\upsilon}}} \v\\
&\times \d
\l[\l(\frac{-\lambda^2_0}{2\sqrt t}z+\lambda_0-i\lambda_0\r)\l(\frac{-\lambda^2_0}{2\sqrt t}z+\lambda_0+i\lambda_0\r)\r]^{-i\upsilon} \v\\
&\d\times e^{-\chi'_{\pm}(\frac{-\lambda^2_0}{2\sqrt t}z-i\lambda_0)+\chi'_{\pm}(i\lambda_0)}e^{-\tilde{\chi}_{\pm}(\frac{-\lambda^2_0}{2\sqrt t}z-i\lambda_0)+\tilde{\chi}_{\pm}(i\lambda_0)},
\end{array}
\ene
with $\chi'_{\pm}(z)$ given by Eq.~(\ref{chii}).\\

Therefore we have the following properties:

\begin{itemize}
\item{} Define $\widehat{M}_{\lambda_0}(x,t,\lambda)$ by
\bee
\widehat{M}_{\lambda_0}(x,t,\lambda)=M^{(4)}(x,t,\lambda)(\delta^0_{\lambda_0})^{\sigma_3},
\ene
where the jump matrix $\widehat{J}_{\lambda_0}(x,t,\lambda)=(\delta^0_{\lambda_0})^{-\hat{\sigma}_3}J^{(4)}(x,t,\lambda)$ is given for $\lambda\in D_\varepsilon (\lambda_0)$ by
\begin{center}
$\widehat{J}_{\lambda_0}(x,t,\lambda)=\left\{
                 \begin{array}{ll}
                 (\delta^0_{\lambda_0})^{-\hat{\sigma}_3}\left(
                 \begin{array}{cc}
           1& -\overline{r_{2r}(\bar{\lambda})}\delta_-^2e^{-t\Phi} \v\\
               0&1 \\
                 \end{array}
                 \right)\left(
                 \begin{array}{cc}
           1& 0 \v\\
               r_{2r}(\lambda)\delta_+^{-2}e^{t\Phi}&1
                 \end{array}
                 \right),       & \lambda \in (3\cap4)\cap D_\varepsilon (\lambda_0), \vspace{0.1in}  \\

         (\delta^0_{\lambda_0})^{-\hat{\sigma}_3} \left(       \begin{array}{cc}
            1& \overline{r_{1r}(\bar{\lambda})}\delta^2e^{-t\Phi} \v\\
               0&1  \end{array}\right)
                 \left( \begin{array}{cc} 1& 0 \v\\
               -r_{1r}(\lambda)\delta^{-2}e^{t\Phi}&1
                 \end{array}
                 \right), & \lambda \in (1\cap2)\cap D_\varepsilon (\lambda_0), \vspace{0.1in}  \\
           \left(
                 \begin{array}{cc}
                  1& -\overline{r_{2a}(\bar{\lambda})}(\delta^1_{\lambda_0})^{2} \v\\
               0&1  \end{array}
                 \right), & \lambda \in (3\cap6)\cap D_\varepsilon (\lambda_0), \vspace{0.1in}  \\
            \left(
                 \begin{array}{cc}         1& 0 \v\\
               r_{2a}(\lambda)(\delta^1_{\lambda_0})^{-2}&1
                 \end{array}
                 \right), & \lambda \in (4\cap8)\cap D_\varepsilon (\lambda_0), \vspace{0.1in}  \\
          \left(
                 \begin{array}{cc}
           1& 0 \v\\
              -(r_{1a}(\lambda)+h_a(\lambda))(\delta^1_{\lambda_0})^{-2}&1
                 \end{array}
                 \right), & \lambda \in (1\cap5)\cap D_\varepsilon (\lambda_0), \vspace{0.1in}  \\
        \left(\begin{array}{cc}
           1& (\overline{r_{1a}(\bar{\lambda})}+\overline{h_a(\bar{\lambda})})(\delta^1_{\lambda_0})^{2} \v\\
            0  &1
                 \end{array}
                 \right), & \lambda \in (2\cap7)\cap D_\varepsilon (\lambda_0),\v\\
                (\delta^0_{\lambda_0})^{-\hat{\sigma}_3}
                 \left(
                 \begin{array}{cc}
           1&0 \v\\
           h_{r}(\lambda)\delta^{-2}e^{t\Phi}&1
                 \end{array}
                 \right), & \lambda \in (5\cap6)\cap D_\varepsilon (\lambda_0),\v\\
                 (\delta^0_{\lambda_0})^{-\hat{\sigma}_3}
                 \left(
                 \begin{array}{cc}
           1& -\overline{h_{r}(\bar{\lambda})}\delta^{2}e^{-t\Phi} \v\\
          0&1   \end{array}
                 \right), & \lambda \in (7\cap8)\cap D_\varepsilon (\lambda_0).
                 \end{array}
                 \right.$
\end{center}

with
\bee\begin{array}{l}
(\delta^1_{\lambda_0})^{-2}r_{2a}(\lambda)-r_{2a}(\lambda_0)z^{2i\upsilon}e^{i\frac{z^2}{2}}\rightarrow 0, \ \ \ t\rightarrow\infty, \v\\
z\rightarrow0\Rightarrow\lambda\rightarrow\lambda_0,\,\, r_{2a}(\lambda)\rightarrow \d\frac{r(\lambda_0)}{1-|r(\lambda_0)|^2},\,\, \,\, r_{1a}(\lambda)+h_{a}(\lambda)\rightarrow r(\lambda_0),
\end{array}
\ene
 combine to Proposition 3.2, we have $\widehat{J}_{\lambda_0}(x,t,z)$ approaches to $J^{X}(x,t,z)$ if $p=r(\lambda_0)$ for $t\rightarrow\infty$ near $z=0$.

Thus we approximate $M^{(4)}$ in the neighborhood $D_\varepsilon (\lambda_0)$ of $\lambda_0$ by $2\times2$ matrix valued function  $M^{\lambda_0}$ of the form
\bee\left\{\begin{array}{l}
M^{\lambda_0}(x,t,\lambda)=(\delta^0_{\lambda_0})^{\sigma_3}M^{X}(r(\lambda_0),z)(\delta^0_{\lambda_0})^{-\sigma_3},\v\\
M^{\lambda_0}(x,t,\lambda)\rightarrow I\,\, {\rm on} \,\, \partial D_\varepsilon (\lambda_0)\,\, {\rm as}\,\,  t\rightarrow\infty,
\end{array}\right.
\ene

\item{} Define $\widehat{M}_{-\lambda_0}(x,t,\lambda)$ by
\bee
\widehat{M}_{-\lambda_0}(x,t,\lambda)=M^{(4)}(x,t,\lambda)(\delta^0_{-\lambda_0})^{\sigma_3},
\ene
where the jump matrix $\widehat{J}_{-\lambda_0}(x,t,\lambda)=(\delta^0_{-\lambda_0})^{-\hat{\sigma}_3}J^{(4)}(x,t,\lambda)$
is given for $\lambda\in D_\varepsilon (-\lambda_0)$ by
\bee\no
\widehat{J}_{-\lambda_0}=\left\{
                 \begin{array}{ll}
                 (\delta^0_{-\lambda_0})^{-\hat{\sigma}_3}\left(
                 \begin{array}{cc}
           1& -\overline{r_{2r}(\bar{\lambda})}\delta_-^2e^{-t\Phi} \v\\
               0&1         \end{array}
                 \right)\left(
                 \begin{array}{cc}      1& 0 \v\\
               r_{2r}(\lambda)\delta_+^{-2}e^{t\Phi}&1
                 \end{array}
                 \right),       & \lambda \in (3\cap4)\cap D_\varepsilon (-\lambda_0), \vspace{0.1in}  \\

         (\delta^0_{-\lambda_0})^{-\hat{\sigma}_3} \left(       \begin{array}{cc}
            1& \overline{r_{1r}(\bar{\lambda})}\delta^2e^{-t\Phi} \v\\
               0&1    \end{array}   \right)
                 \left(
                 \begin{array}{cc}   1& 0\v\\
               -r_{1r}(\lambda)\delta^{-2}e^{t\Phi}&1
                 \end{array}
                 \right), & \lambda \in (1\cap2)\cap D_\varepsilon (-\lambda_0), \vspace{0.1in}  \\
           \left(
                 \begin{array}{cc}
                  1& -\overline{r_{2a}(\bar{\lambda})}(\delta^1_{-\lambda_0})^{2} \v\\
               0&1   \end{array}
                 \right), & \lambda \in (3\cap6)\cap D_\varepsilon (-\lambda_0), \vspace{0.1in}  \\
            \left(
                 \begin{array}{cc}    1& 0 \v\\
               r_{2a}(\lambda)(\delta^1_{-\lambda_0})^{-2}&1
                 \end{array}
                 \right), & \lambda \in (4\cap8)\cap D_\varepsilon (-\lambda_0), \vspace{0.1in}  \\
          \left(
                 \begin{array}{cc}      1& 0 \v\\
              -(r_{1a}(\lambda)+h_a(\lambda))(\delta^1_{-\lambda_0})^{-2}&1
                 \end{array}
                 \right), & \lambda \in (1\cap5)\cap D_\varepsilon (-\lambda_0), \vspace{0.1in}  \\
        \left(
                 \begin{array}{cc}
           1& (\overline{r_{1a}(\bar{\lambda})}+\overline{h_a(\bar{\lambda})})(\delta^1_{-\lambda_0})^{2} \v\\
            0  &1               \end{array}
                 \right), & \lambda \in (2\cap7)\cap D_\varepsilon (-\lambda_0),\v\\
                (\delta^0_{-\lambda_0})^{-\hat{\sigma}_3}
                 \left(
                 \begin{array}{cc}
           1&0 \v\\
           h_{r}(\lambda)\delta^{-2}e^{t\Phi}&1
                 \end{array}
                 \right), & \lambda \in (5\cap6)\cap D_\varepsilon (-\lambda_0), \v\\
                 (\delta^0_{-\lambda_0})^{-\hat{\sigma}_3}
                 \left(
                 \begin{array}{cc}
           1&-\overline{h_{r}(\bar{\lambda})}\delta^{2}e^{-t\Phi} \v\\
           0&1          \end{array}
                 \right), & \lambda \in (7\cap8)\cap D_\varepsilon (-\lambda_0).
                 \end{array}
                 \right.
\ene

with
\bee
\begin{array}{l}
(\delta^1_{-\lambda_0})^{-2}r_{2a}(\lambda)-r_{2a}(-\lambda_0)z^{2i\upsilon}e^{i\frac{z^2}{2}}\rightarrow 0, \ \ \ t\rightarrow\infty,\v\\
z\rightarrow0\Rightarrow\lambda\rightarrow-\lambda_0,\,\, r_{2a}(\lambda)\rightarrow \d\frac{r(-\lambda_0)}{1-|r(-\lambda_0)|^2}, \,\, \,\, r_{1a}(\lambda)+h_{a}(\lambda)\rightarrow r(-\lambda_0),
\end{array}
\ene
combine to Proposition 3.2, we know $\widehat{J}_{-\lambda_0}(x,t,z)$ tends to $J^{X}(x,t,z)$ if $p=r(-\lambda_0)$ for $t\rightarrow\infty$ near $z=0$.

Therefore we approximate $M^{(4)}$ in the neighborhood $D_\varepsilon (-\lambda_0)$ of $-\lambda_0$ by $2\times2$ matrix valued function  $M^{-\lambda_0}$ of the form
\bee\l\{\begin{array}{l}
M^{-\lambda_0}(x,t,\lambda)=(\delta^0_{-\lambda_0})^{\sigma_3}M^{X}(r(-\lambda_0),z)(\delta^0_{-\lambda_0})^{-\sigma_3},\v\\
M^{-\lambda_0}(x,t,\lambda)\rightarrow I\,\, {\rm on} \,\, \partial D_\varepsilon (-\lambda_0)\,\, {\rm as}\,\ t\rightarrow\infty.
\end{array}\r.
\ene

\item{} Define $\widehat{M}_{i\lambda_0}(x,t,\lambda)$ by
\bee
\widehat{M}_{i\lambda_0}(x,t,\lambda)=M^{(4)}(x,t,\lambda)(\delta^0_{i\lambda_0})^{\sigma_3},
\ene
where the jump matrix $\widehat{J}_{i\lambda_0}(x,t,\lambda)=(\delta^0_{i\lambda_0})^{-\hat{\sigma}_3}J^{(4)}(x,t,\lambda)$
is given for $\lambda\in D_\varepsilon (i\lambda_0)$ by
\bee\no
\widehat{J}_{i\lambda_0}(x,t,\lambda)=\left\{
                 \begin{array}{ll}
                 (\delta^0_{i\lambda_0})^{-\hat{\sigma}_3}\left(
                 \begin{array}{cc}
           1& -\overline{r_{2r}(\bar{\lambda})}\delta_-^2e^{-t\Phi} \v\\
               0&1     \end{array}
                 \right)\left(
                 \begin{array}{cc}   1& 0 \v\\
               r_{2r}(\lambda)\delta_+^{-2}e^{t\Phi}&1
                 \end{array}
                 \right),
                 & \lambda \in (3\cap4)\cap D_\varepsilon (i\lambda_0), \vspace{0.1in}  \\
(\delta^0_{i\lambda_0})^{-\hat{\sigma}_3}
          \left(
                 \begin{array}{cc}
            1& \overline{r_{1r}(\bar{\lambda})}\delta^2e^{-t\Phi} \v\\
               0&1            \end{array}
                 \right)
                 \left(
                 \begin{array}{cc}   1& 0\v\\
               -r_{1r}(\lambda)\delta^{-2}e^{t\Phi}&1
                 \end{array}
                 \right), & \lambda \in (1\cap2)\cap D_\varepsilon (i\lambda_0), \vspace{0.1in}  \\
           \left(
                 \begin{array}{cc}
                  1& -\overline{r_{2,a}(\bar{\lambda})}(\delta^1_{i\lambda_0})^{2} \v\\
               0&1  \end{array}
                 \right), & \lambda \in (3\cap6)\cap D_\varepsilon (i\lambda_0), \vspace{0.1in}  \\
            \left(
                 \begin{array}{cc}   1& 0 \v\\
               r_{2a}(\lambda)(\delta^1_{i\lambda_0})^{-2}&1
                 \end{array}
                 \right), & \lambda \in (4\cap8)\cap D_\varepsilon (i\lambda_0), \vspace{0.1in}  \\
          \left(
                 \begin{array}{cc}   1& 0 \v\\
              -(r_{1a}(\lambda)+h_a(\lambda))(\delta^1_{i\lambda_0})^{-2}&1
                 \end{array}
                 \right), & \lambda \in (1\cap5)\cap D_\varepsilon (i\lambda_0), \vspace{0.1in}  \\
        \left(
                 \begin{array}{cc}
           1& (\overline{r_{1a}(\bar{\lambda})}+\overline{h_a(\bar{\lambda})})(\delta^1_{i\lambda_0})^{2} \v\\
            0  &1             \end{array}
                 \right), & \lambda \in (2\cap7)\cap D_\varepsilon (i\lambda_0),\v\\
                  (\delta^0_{i\lambda_0})^{-\hat{\sigma}_3}
                 \left(
                 \begin{array}{cc}      1&0 \v\\
           h_{r}(\lambda)\delta^{-2}e^{t\Phi}&1
                 \end{array}
                 \right), & \lambda \in (5\cap6)\cap D_\varepsilon (i\lambda_0),\v\\
                 (\delta^0_{i\lambda_0})^{-\hat{\sigma}_3}
                 \left(
                 \begin{array}{cc}
           1&-\overline{h_{r}(\bar{\lambda})}\delta^{2}e^{-t\Phi} \v\\
           0&1
                 \end{array}
                 \right), & \lambda \in (7\cap8)\cap D_\varepsilon (i\lambda_0).
                 \end{array}
                 \right.
\ene

with
\bee\begin{array}{l}
(\delta^1_{i\lambda_0})^{-2}r_{2a}(\lambda)-r_{2a}(i\lambda_0)z^{-2i\tilde{\upsilon}}e^{i\frac{z^2}{2}}\rightarrow 0, \ \ \ t\rightarrow\infty,\v\\
z\rightarrow0\Rightarrow\lambda\rightarrow i\lambda_0,\,\, r_{2a}(\lambda)\rightarrow \d\frac{r(i\lambda_0)}{1-r(i\lambda_0)\overline{r(-i\lambda_0)}},\,\,\,\, r_{1a}(\lambda)+h_{a}(\lambda)\rightarrow r(i\lambda_0),
\end{array}
\ene
combine to Proposition 3.2, we have $\widehat{J}_{i\lambda_0}(x,t,z)$ tends to $J^{Y}(x,t,z)$ if $p=r(i\lambda_0)$ for $t\rightarrow\infty$ near $z=0$.

Therefore we approximate $M^{(4)}$ in the neighborhood $D_\varepsilon (i\lambda_0)$ of $i\lambda_0$ by $2\times2$ matrix valued function  $M^{i\lambda_0}$ of the form
\bee\l\{\begin{array}{l}
M^{i\lambda_0}(x,t,\lambda)=(\delta^0_{i\lambda_0})^{\sigma_3}M^{Y}(r(i\lambda_0),z)(\delta^0_{i\lambda_0})^{-\sigma_3},\v\\
M^{i\lambda_0}(x,t,\lambda)\rightarrow I\,\, {\rm on}\,\, \partial D_\varepsilon (i\lambda_0)\,\, {\rm as} \,\, t\rightarrow\infty.
\end{array}\r.
\ene

\item{} Define $\widehat{M}_{-i\lambda_0}(x,t,\lambda)$ by
\bee
\widehat{M}_{-i\lambda_0}(x,t,\lambda)=M^{(4)}(x,t,\lambda)(\delta^0_{-i\lambda_0})^{\sigma_3},
\ene
where the jump matrix $\widehat{J}_{-i\lambda_0}(x,t,\lambda)=(\delta^0_{-i\lambda_0})^{-\hat{\sigma}_3}J^{(4)}(x,t,\lambda)$
is given for $\lambda\in D_\varepsilon (-i\lambda_0)$ by
\bee\no
\widehat{J}_{-i\lambda_0}=\left\{
                 \begin{array}{ll}
                (\delta^0_{-i\lambda_0})^{-\hat{\sigma}_3} \left(
                 \begin{array}{cc}
           1& -\overline{r_{2r}(\bar{\lambda})}\delta_-^2e^{-t\Phi} v\\
               0&1                  \end{array}
                 \right)\left(
                 \begin{array}{cc}          1& 0\v\\
               r_{2r}(\lambda)\delta_+^{-2}e^{t\Phi}&1
                 \end{array}
                 \right),
                 & \lambda \in (3\cap4)\cap D_\varepsilon (-i\lambda_0), \vspace{0.1in}  \\
(\delta^0_{-i\lambda_0})^{-\hat{\sigma}_3}
          \left(
                 \begin{array}{cc}
            1& \overline{r_{1r}(\bar{\lambda})}\delta^2e^{-t\Phi} \v\\
               0&1  \end{array}  \right)
                 \left(
                 \begin{array}{cc}  1& 0 \v\\
               -r_{1r}(\lambda)\delta^{-2}e^{t\Phi}&1
                 \end{array}
                 \right), & \lambda \in (1\cap2)\cap D_\varepsilon (-i\lambda_0), \vspace{0.1in}  \\
           \left(
                 \begin{array}{cc}
                  1& -\overline{r_{2a}(\bar{\lambda})}(\delta^1_{-i\lambda_0})^{2} \v\\
               0&1   \end{array}
                 \right) & \lambda \in (3\cap6)\cap D_\varepsilon (-i\lambda_0), \vspace{0.1in}  \\
            \left(
                 \begin{array}{cc}      1& 0 \v\\
               r_{2a}(\lambda)(\delta^1_{-i\lambda_0})^{-2}&1
                 \end{array}
                 \right), & \lambda \in (4\cap8)\cap D_\varepsilon (-i\lambda_0), \vspace{0.1in}  \\
          \left(
                 \begin{array}{cc}       1& 0 \v\\
              -(r_{1a}(\lambda)+h_a(\lambda))(\delta^1_{-i\lambda_0})^{-2}&1
                 \end{array}
                 \right), & \lambda \in (1\cap5)\cap D_\varepsilon (-i\lambda_0), \vspace{0.1in}  \\
        \left(
                 \begin{array}{cc}
           1& (\overline{r_{1a}(\bar{\lambda})}+\overline{h_a(\bar{\lambda})})(\delta^1_{-i\lambda_0})^{2} \v\\
            0  &1
                 \end{array}
                 \right), & \lambda \in (2\cap7)\cap D_\varepsilon (-i\lambda_0),\v\\
                   (\delta^0_{-i\lambda_0})^{-\hat{\sigma}_3}
                 \left(
                 \begin{array}{cc} 1&0 \v\\
           h_{r}(\lambda)\delta^{-2}e^{t\Phi}&1
                 \end{array}
                 \right), & \lambda \in (5\cap6)\cap D_\varepsilon (-i\lambda_0),\v\\
                 (\delta^0_{-i\lambda_0})^{-\hat{\sigma}_3}
                 \left(
                 \begin{array}{cc}
           1&-\overline{h_{r}(\bar{\lambda})}\delta^{2}e^{-t\Phi} \v\\
           0&1    \end{array}
                 \right), & \lambda \in (7\cap8)\cap D_\varepsilon (-i\lambda_0).
                 \end{array}
                 \right.
\ene

with
\bee\begin{array}{l}
(\delta^1_{-i\lambda_0})^{-2}r_{2a}(\lambda)-r_{2a}(-i\lambda_0)z^{-2i\tilde{\upsilon}}e^{i\frac{z^2}{2}}\rightarrow 0, \ \ \ t\rightarrow\infty,\v\\
z\rightarrow0\Rightarrow\lambda\rightarrow -i\lambda_0,\,\, r_{2a}(\lambda)\rightarrow \d\frac{r(-i\lambda_0)}{1-r(-i\lambda_0)\overline{r(i\lambda_0)}},\,\,\,\, r_{1a}(\lambda)+h_{a}(\lambda)\rightarrow r(-i\lambda_0),
\end{array}
\ene
combine to Proposition 3.2, we have $\widehat{J}_{-i\lambda_0}(x,t,z)$ tends to $J^{Y}(x,t,z)$ if $p=r(-i\lambda_0)$ for $t\rightarrow\infty$ near $z=0$.

Therefore we approximate $M^{(4)}$ in the neighborhood $D_\varepsilon (-i\lambda_0)$ of $-i\lambda_0$ by $2\times2$ matrix valued function  $M^{-i\lambda_0}$ of the form
\bee\l\{\begin{array}{l}
M^{-i\lambda_0}(x,t,\lambda)=(\delta^0_{-i\lambda_0})^{\sigma_3}M^{Y}(r(-i\lambda_0),z)(\delta^0_{-i\lambda_0})^{-\sigma_3},\v\\
M^{-i\lambda_0}(x,t,\lambda)\rightarrow I\,\, {\rm on}\,\, \partial D_\varepsilon (-i\lambda_0)\,\, {\rm as}
\,\,t\rightarrow\infty.
\end{array}\r.
\ene
\end{itemize}

\begin{figure}[!t]
\begin{center}
\vspace{0.05in}
{\scalebox{0.526}[0.42]{\includegraphics{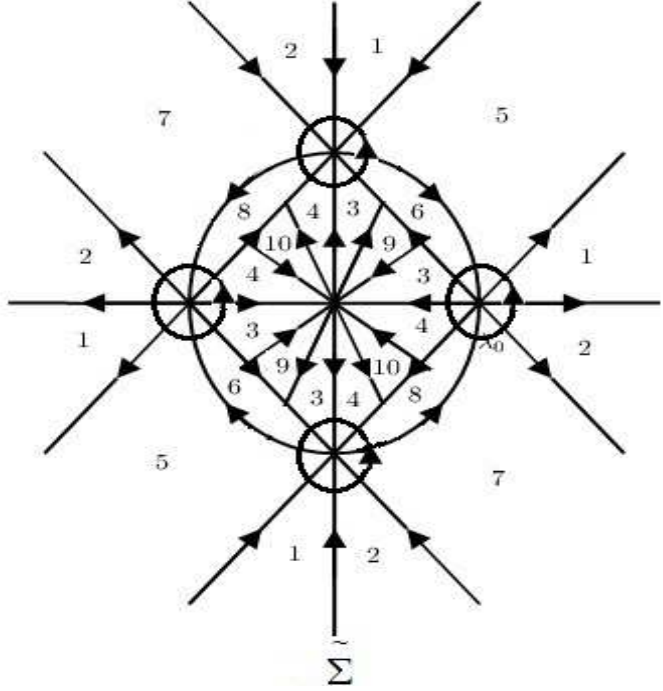}}}
\end{center}
\vspace{-0.25in}\caption{The jump contour $\widetilde{\Sigma}$ of the new modification.}
\label{x-region}
\end{figure}

\noindent \textbf{Proposition 4.1}
For each $\varsigma\in(0,N)$ and $t>0$,\, the jump matrix $J^{j}$ of $M^{j}_+=M^{j}_-J^{j}$ satisfies
\bee
||J^{(4)}-J^{j}||_{L^1\cup L^2 \cup L^\infty(\Sigma^{(2)}\cap D_\varepsilon(j))}\leq c\frac{\ln t}{\sqrt{t}},\quad j=\lambda_0,-\lambda_0,i\lambda_0,-i\lambda_0.
\ene

Notice that the proposition is a consequence of Proposition 3.1 in Ref.~\cite{xu15}.

\section{Main results}

Define the approximate solution $M^{a}$ by
\bee
M^{a}=\left\{
                 \begin{array}{ll}

                 M^{\lambda_0}, & \lambda \in  D_\varepsilon (\lambda_0), \vspace{0.1in}  \\
                 M^{-\lambda_0}, & \lambda \in  D_\varepsilon (-\lambda_0), \vspace{0.1in}  \\
                 M^{i\lambda_0}, & \lambda \in  D_\varepsilon (i\lambda_0), \vspace{0.1in}  \\
                 M^{-i\lambda_0}, & \lambda \in  D_\varepsilon (-i\lambda_0), \vspace{0.1in}  \\
        I, & {\rm elsewhere}.
                 \end{array}
                 \right.
\ene
We will show that the solution $\widetilde{M}$ defined by
\bee
\widetilde{M}=M^{(4)}(M^{a})^{-1}
\ene
is small for $t\rightarrow\infty$.

The RH problem $(\widetilde{M},\widetilde{J}(x,t,\lambda),\widetilde{\Sigma})$ with $\widetilde{\Sigma}=\Sigma^{(2)}\cup\partial D_\varepsilon (\lambda_0)\cup\partial D_\varepsilon (-\lambda_0)\cup\partial D_\varepsilon (i\lambda_0)\cup\partial D_\varepsilon (-i\lambda_0)$ (see Figure 7) is given by
\begin{itemize}
\item{} $\widetilde{M}(x,t,\lambda)$  is in general a meromorphic function in $\lambda\in\mathbb{C}\backslash\  \widetilde{\Sigma}$;

\item{} $\widetilde{M}_{+}(x,t,\lambda)=\widetilde{M}_{-}(x,t,\lambda)\widetilde{J}(x,t,\lambda)$   for $\lambda\in \widetilde{\Sigma}$ ;

\item{}$\widetilde{M}(x,t,\lambda)=I+O(\frac{1}{\lambda})$ as $\lambda\rightarrow\infty.$
\end{itemize}
where the jump matrix $\widetilde{J}(x,t,\lambda)=(M^{a})_{-}J^{(4)}(M^{a})_{+}^{-1}$ is
\bee
\widetilde{J}(x,t,\lambda)=\left\{
                 \begin{array}{ll}
                 J^{(4)}, & \lambda \in  \widetilde{\Sigma}\setminus (\overline{D_\varepsilon (\lambda_0)}\cup\overline{D_\varepsilon (-\lambda_0)}\cup\overline{D_\varepsilon (i\lambda_0)}\cup\overline{D_\varepsilon (-i\lambda_0)}), \vspace{0.1in}  \\
                 (M^{a})_+^{-1},& \lambda \in  \partial D_\varepsilon (\lambda_0)\cup\partial D_\varepsilon (-\lambda_0)\cup\partial D_\varepsilon (i\lambda_0)\cup\partial D_\varepsilon (-i\lambda_0), \vspace{0.1in}  \\
                (M^{a})_{-}J^{(4)}(M^{a})_{+}^{-1}, & \lambda \in  \widetilde{\Sigma}\cap (D_\varepsilon (\lambda_0)\cup D_\varepsilon (-\lambda_0)\cup D_\varepsilon (i\lambda_0)\cup D_\varepsilon (-i\lambda_0)). \vspace{0.1in}  \\
                 \end{array}
                 \right.
\ene

\v
\noindent {\it 5.1.\, Long-time asymptotics of $\widetilde{M}(x,t,\lambda)$} \v

Define
\bee
 \Sigma'=\widetilde{\Sigma}\setminus [\partial D_\varepsilon(\lambda_0)\cup \partial D_\varepsilon(-\lambda_0)\cup \partial D_\varepsilon(i\lambda_0)\cup \partial D_\varepsilon(-i\lambda_0)\cup \mathcal{X}_{\lambda_0}^\varepsilon\cup \mathcal{X}_{-\lambda_0}^\varepsilon\cup \mathcal{X}_{i\lambda_0}^\varepsilon \cup \mathcal{X}_{-i\lambda_0}^\varepsilon],
 \ene
where $\mathcal{X}_{\lambda_0}^\varepsilon=\Sigma^{(2)}\cap D_\varepsilon(\lambda_0)$ stands for denote the part of X that lies in the disk $D_\varepsilon(\lambda_0)$, and $\mathcal{X}_{-\lambda_0}^\varepsilon$ , $\mathcal{X}_{i\lambda_0}^\varepsilon$, $\mathcal{X}_{-i\lambda_0}^\varepsilon$ have the similar definitions.

\v\noindent \textbf{Proposition 5.1} The function $\widetilde{w}(x,t,\lambda):=\widetilde{J}(x,t,\lambda)-I $ satisfies
\bee
\label{w}
\begin{array}{l}
||\widetilde{w}(x,t,\lambda)||_{(L^1 \cup L^2 \cup L^\infty)(\Sigma')}=O(Ct^{-\frac{3}{2}}),\ \ t\rightarrow\infty ,\, \varsigma\in (0,N),\v\\
||\widetilde{w}(x,t,\lambda)||_{L^1(\mathcal{X}_{\lambda_0}^{\varepsilon})}=O(\frac{\ln t}{t}),\ \ t\rightarrow\infty ,\, \varsigma\in (0,N),\v\\
||\widetilde{w}(x,t,\lambda)||_{L^1(\mathcal{X}_{-\lambda_0}^{\varepsilon})}=O(\frac{\ln t}{t}),\ \ t\rightarrow\infty ,\, \varsigma\in (0,N),\v\\
||\widetilde{w}(x,t,\lambda)||_{L^1(\mathcal{X}_{i\lambda_0}^{\varepsilon})}=O(\frac{\ln t}{t}),\ \ t\rightarrow\infty ,\, \varsigma\in (0,N),\v\\
||\widetilde{w}(x,t,\lambda)||_{L^1(\mathcal{X}_{-i\lambda_0}^{\varepsilon})}=O(\frac{\ln t}{t}),\ \ t\rightarrow\infty ,\, \varsigma\in (0,N).
\end{array}
\ene
where the error term is uniform with respect to $(\varsigma,\lambda)$ in the given ranges.

\v\noindent {\bf Proof.}\,  For $\lambda\in\Sigma'$, \, $\widetilde{w}=\widetilde{J}-I=J^{(4)}-I$ involves the small remainders $h_r$, $r_{1r}$, $r_{2r}$. Moreover, according to Proposition 3.2, we can show that the first one in system (~\ref{w}) holds. For $\lambda\in\mathcal{X}_{\lambda_0}^{\varepsilon}\cup\mathcal{X}_{-\lambda_0}^{\varepsilon}\cup
\mathcal{X}_{i\lambda_0}^{\varepsilon}\cup\mathcal{X}_{-i\lambda_0}^{\varepsilon}$, $\widetilde{w}=\widetilde{J}-I=(M^a)_-J^{(4)}(M^a)_+^{-1}-I$,
we can show that the other equations in system (~\ref{w}) also hold as a consequence of the define of $M^a$ and Proposition 4.1.

\vspace{0.1in} \noindent  \textbf{Proposition 5.2}~\cite{le15} Let $\mathcal{\widetilde{C}}_{\widetilde{w}}$ denote the operator associated with $\widetilde{\Sigma}$, i.e., $\mathcal{\widetilde{C}}_{\widetilde{w}}:\, L^2(\widetilde{\Sigma})+L^{\infty}(\widetilde{\Sigma}) \to L^2(\widetilde{\Sigma})$ with $\mathcal{\widetilde{C}}_{\widetilde{w}}f=\mathcal{\widetilde{C}}_{-}(f\widetilde{w})$, where $(\mathcal{\widetilde{C}}_{-})f(z)=\frac{1}{2\pi i}\int_{\widetilde{\Sigma}}\frac{f(s)}{s-z_{-}}ds\, z\in \mathbb{C}\setminus \widetilde{\Sigma}$. Then there exists a $T>0$ such that $I-\mathcal{\widetilde{C}}_{\widetilde{w}}\in \mathcal{B}(L^2(\widetilde{\Sigma}))$
is invertible for all $(\varsigma,t)\in (0,N)\times(0,\infty)$ with $t>T$. Moreover, the function $\hat{\mu}(\varsigma,t,\lambda)=I+(I-\mathcal{\widetilde{C}}_{\widetilde{w}})^{-1}\mathcal{\widetilde{C}}_{\widetilde{w}}I \in I+L^2(\widetilde{\Sigma})$ satisfies
\begin{eqnarray}\label{5}
\|\widetilde{\mu}(\varsigma,t,\lambda)-I\|_{L^2(\widetilde{\sum})}=O(Ct^{-\frac{1}{2}}),\ \ t\rightarrow\infty,\, \varsigma\in (0,N)
\end{eqnarray}
where  error terms are uniform with respect to $\varsigma=\frac{x}{t}$.\\

\noindent \textbf{Proposition 5.3}~\cite{le15} The RH problem $(\widetilde{M},\widetilde{J}(x,t,\lambda),\widetilde{\Sigma})$ admits the unique solution given by
\bee
\widetilde{M}(x,t,\lambda)=I+\frac{1}{2\pi i}\int_{\widetilde{\Sigma}}
\frac{\widetilde{\mu}(\varsigma,t,s)\widetilde{w}(\varsigma,t,s)}{s-\lambda}ds,
\ene
for $t>T$. Moreover, for each point $(\zeta, t)\in (0,N)\times (0, \infty)$ and $t>T$, the nontangential limit of $\lambda(\widetilde{M}(\varsigma,t,\lambda)-I)$ is defined by
\begin{center}
$\lim\limits_{\lambda\rightarrow\infty}\lambda(\widetilde{M}(\varsigma,t,\lambda)-I)=\d\frac{i}{2\pi}\int_{\widetilde{\Sigma}}
\widetilde{\mu}(\varsigma,t,\lambda)\widetilde{w}(\varsigma,t,\lambda)d\lambda.$ \\
\end{center}

\vspace{0.1in}

By using the expressions of $M^X$ and $M^Y$, we can get the expressions of $(M^{\lambda_0})^{-1}(\varsigma,t,\lambda)$,\, $(M^{-\lambda_0})^{-1}(\varsigma,t,\lambda)$,\, $(M^{i\lambda_0})^{-1}(\varsigma,t,\lambda)$,\,
$(M^{-i\lambda_0})^{-1}(\varsigma,t,\lambda)$ as
\bee\begin{array}{l}
(M^{\lambda_0})^{-1}(x,t,\lambda)=(\delta^0_{\lambda_0})^{\sigma_3}(M^{X}(r(\lambda_0),z))^{-1}(\delta^0_{\lambda_0})^{-\sigma_3}, \v\\
(M^{-\lambda_0})^{-1}(x,t,\lambda)=(\delta^0_{-\lambda_0})^{\sigma_3}(M^{X}(r(-\lambda_0),z))^{-1}(\delta^0_{-\lambda_0})^{-\sigma_3}, \v\\
(M^{i\lambda_0})^{-1}(x,t,\lambda)=(\delta^0_{i\lambda_0})^{\sigma_3}(M^{Y}(r(i\lambda_0),z))^{-1}(\delta^0_{i\lambda_0})^{-\sigma_3}, \v\\
(M^{-i\lambda_0})^{-1}(x,t,\lambda)=(\delta^0_{-i\lambda_0})^{\sigma_3}(M^{Y}(r(-i\lambda_0),z))^{-1}(\delta^0_{-i\lambda_0})^{-\sigma_3}, \v\\
\end{array}
\ene

Thus we have

{\it Case 1.} For the variable $z= \frac{2\sqrt t}{\lambda^2_0}(\lambda-\lambda_0)$, thus
\bee
M^X(r(\lambda_0),z)=I-\d\frac{M_{1}^X(r(\lambda_0))}{\frac{2\sqrt t}{\lambda^2_0}(\lambda-\lambda_0)}+O\l(\frac{r(\lambda_0)}{t}\r), \ \ z\rightarrow\infty,
\ene
with
\bee \no
M_{1}^X(r(\lambda_0))=i\left(
                 \begin{array}{cc}
           0& \beta^X(r(\lambda_0)) \v\\
              \overline{\beta^X(r(\lambda_0))} & 0
                 \end{array}
                 \right),
\ene
Thus
\bee\begin{array}{rl}\label{3}
(M^{\lambda_0})^{-1}(\varsigma,t,\lambda)=&(\delta^0_{\lambda_0})^{\sigma_3}(M^X(\varsigma))^{-1}(\delta^0_{\lambda_0})^{-\sigma_3} \vspace{0.15in}\\
=&\d I+\frac{(\delta^0_{\lambda_0})^{\hat{\sigma}_3}M_{1}^X(r(\lambda_0))}{\frac{2\sqrt t}{\lambda^2_0}(\lambda-\lambda_0)}+O\l(\frac{r(\lambda_0)}{t}\r), \, \, t\rightarrow\infty,\,\,\lambda\in \partial D_\varepsilon (\lambda_0),
\end{array}\ene
which and $|M_{1}^X(r(\lambda_0))|\leq c|r(\lambda_0)|$ generate
\bee
||(M^{\lambda_0})^{-1}-I||_{L^1\cup L^2 \cup L^\infty(\partial D_\varepsilon(\lambda_0))}
=O\l(r(\lambda_0)t^{-\frac{1}{2}}\r).
\ene

{\it Case 2.} For the variable $z= \frac{2\sqrt t}{\lambda^2_0}(\lambda+\lambda_0)$, we have
\bee
M^X(r(-\lambda_0),z)=I-\frac{M_{1}^X(r(-\lambda_0))}{\frac{2\sqrt t}{\lambda^2_0}(\lambda+\lambda_0)}+O\l(\frac{r(-\lambda_0)}{t}\r), \ \ z\rightarrow\infty.
\ene
with
\bee\no
M_{1}^X(r(-\lambda_0))=i\left(
                 \begin{array}{cc}
           0& \beta^X(r(-\lambda_0)) \v\\
              \overline{\beta^X(r(-\lambda_0))} & 0
                 \end{array}
                 \right),
\ene
Thus
\bee\begin{array}{rl}\label{4}
(M^{-\lambda_0})^{-1}(\varsigma,t,\lambda)=&(\delta^0_{-\lambda_0})^{\sigma_3}(M^X(\varsigma))^{-1}(\delta^0_{-\lambda_0})^{-\sigma_3} \vspace{0.15in}\\
=&\d I+\frac{(\delta^0_{-\lambda_0})^{\hat{\sigma}_3}M_{1}^X(r(-\lambda_0))}{\frac{2\sqrt t}{\lambda^2_0}(\lambda+\lambda_0)}+O\l(\frac{r(-\lambda_0)}{t}\r), \ \ t\rightarrow\infty,\,\,\lambda\in \partial D_\varepsilon (-\lambda_0),
\end{array}\ene
which and $|M_{1}^X(r(-\lambda_0))|\leq c|r(-\lambda_0)|$ lead to
\bee
||(M^{-\lambda_0})^{-1}-I||_{L^1\cup L^2 \cup L^\infty(\partial D_\varepsilon(-\lambda_0))}
=O\l(r(-\lambda_0)t^{-\frac{1}{2}}\r).
\ene

{\it Case 3.} For the variable $z= -\frac{2\sqrt t}{\lambda^2_0}(\lambda-i\lambda_0)$, thus
\bee
M^Y(r(i\lambda_0),z)=I+\frac{M_{1}^Y(r(i\lambda_0))}{\frac{2\sqrt t}{\lambda^2_0}(\lambda-i\lambda_0)}+O\l(\frac{r(i\lambda_0)}{t}\r), \ \ z\rightarrow\infty.
\ene
with
\bee\no
M_{1}^Y(r(i\lambda_0))=i\left(
                 \begin{array}{cc}
           0& \beta^Y(r(i\lambda_0)) \v\\
              -\overline{\beta^Y(r(\overline{i\lambda_0}))} & 0
                 \end{array}
                 \right),
\ene
Thus
\bee\begin{array}{rl}\label{55}
(M^{i\lambda_0})^{-1}(\varsigma,t,\lambda)=&(\delta^0_{i\lambda_0})^{\sigma_3}(M^Y(r(i\lambda_0)))^{-1}(\delta^0_{i\lambda_0})^{-\sigma_3} \vspace{0.15in}\\
=&\d I-\frac{(\delta^0_{i\lambda_0})^{\hat{\sigma}_3}M_{1}^Y(r(i\lambda_0))}{\frac{2\sqrt t}{\lambda^2_0}(\lambda-i\lambda_0)}+O\l(\frac{r(i\lambda_0)}{t}\r), \ \ t\rightarrow\infty,\,\,\lambda\in \partial D_\varepsilon (i\lambda_0),
\end{array}\ene
which and $|M_{1}^X(r(i\lambda_0))|\leq c|r(i\lambda_0)|$ generate
\bee
||(M^{i\lambda_0})^{-1}-I||_{L^1\cup L^2 \cup L^\infty(\partial D_\varepsilon(i\lambda_0))}
=O\l(r(i\lambda_0)t^{-\frac{1}{2}}\r).
\ene

{\it Case 4.} For the variable $z= -\frac{2\sqrt t}{\lambda^2_0}(\lambda+i\lambda_0)$, thus
\bee
M^Y(r(-i\lambda_0),z)=I+\frac{M_{1}^Y(r(-i\lambda_0))}{\frac{2\sqrt t}{\lambda^2_0}(\lambda+i\lambda_0)}+O\l(\frac{r(-i\lambda_0)}{t}\r), \ \ z\rightarrow\infty.
\ene
with \bee\no
M_{1}^Y(r(-i\lambda_0))=i\left(
                 \begin{array}{cc}
           0& \beta^Y(r(-i\lambda_0)) \v\\
              -\overline{\beta^Y(r(\overline{-i\lambda_0}))} & 0
                 \end{array}
                 \right),
\ene
Thus we get
\bee\begin{array}{rl}\label{6}
(M^{-i\lambda_0})^{-1}(\varsigma,t,\lambda)=&(\delta^0_{-i\lambda_0})^{\sigma_3}(M^Y(r(-i\lambda_0)))^{-1}(\delta^0_{-i\lambda_0})^{-\sigma_3} \vspace{0.15in}\\
=&\d I-\frac{(\delta^0_{-i\lambda_0})^{\hat{\sigma}_3}M_{1}^Y(r(-i\lambda_0))}{\frac{2\sqrt t}{\lambda^2_0}(\lambda+i\lambda_0)}+O\l(\frac{r(-i\lambda_0)}{t}\r), \ \ t\rightarrow\infty,\,\,\lambda\in \partial D_\varepsilon (-i\lambda_0),
\end{array}\ene
which and $|M_{1}^X(r(-i\lambda_0))|\leq c|r(-i\lambda_0)|$ lead to
\bee
||(M^{-i\lambda_0})^{-1}-I||_{L^1\cup L^2 \cup L^\infty(\partial D_\varepsilon(-i\lambda_0))}=O\l(r(-i\lambda_0)t^{-\frac{1}{2}}\r).
\ene

By using Proposition 5.1 and Eqs.~(\ref{5}), (\ref{3}), (\ref{4}), (\ref{55}), (\ref{6}), and the H\"{o}lder inequation, we can obtain
\bee \begin{array}{rl}\label{666}
\d\int_{|\lambda-\lambda_{0}|=\epsilon}
\widetilde{\mu}(\varsigma,t,\lambda)((M^{a})_+^{-1}(\varsigma,t,\lambda)-I)\mathrm{d}\lambda
=&\d\int_{|\lambda-\lambda_{0}|=\epsilon}((M^{a})_+^{-1}(\varsigma,t,\lambda)-I)\mathrm{d}\lambda  \vspace{0.15in} \\
&\d\quad +\int_{|\lambda-\lambda_{0}|=\epsilon}
(\widetilde{\mu}(\varsigma,t,\lambda)-I)((M^{a})_+^{-1}(\varsigma,t,\lambda)-I)\mathrm{d}\lambda \vspace{0.15in}\\
=&\d 2\pi i\frac{(\delta^0_{\lambda_0})^{\hat{\sigma}_3}M_{1}^X(r(\lambda_0))}{\frac{2\sqrt t}{\lambda^2_0}}+O\l(\frac{r(\lambda_0)}{t}\r),\quad t\rightarrow\infty,\,
\end{array}
\ene

\bee \begin{array}{rl}\label{666}
\d\int_{|\lambda+\lambda_{0}|=\epsilon}
\widetilde{\mu}(\varsigma,t,\lambda)((M^{a})_+^{-1}(\varsigma,t,\lambda)-I)\mathrm{d}\lambda
=&\d\int_{|\lambda+\lambda_{0}|=\epsilon}((M^{a})_+^{-1}(\varsigma,t,\lambda)-I)\mathrm{d}\lambda  \vspace{0.15in} \\
&\d\quad +\int_{|\lambda+\lambda_{0}|=\epsilon}
(\widetilde{\mu}(\varsigma,t,\lambda)-I)((M^{a})_+^{-1}(\varsigma,t,\lambda)-I)\mathrm{d}\lambda \vspace{0.15in}\\
=&\d 2\pi i\frac{(\delta^0_{-\lambda_0})^{\hat{\sigma}_3}M_{1}^X(r(-\lambda_0))}{\frac{2\sqrt t}{\lambda^2_0}}+O\l(\frac{r(-\lambda_0)}{t}\r),\quad t\rightarrow\infty,\,
\end{array}
\ene
\bee \begin{array}{rl}\label{666}
\d\int_{|\lambda-i\lambda_{0}|=\epsilon}
\widetilde{\mu}(\varsigma,t,\lambda)((M^{a})_+^{-1}(\varsigma,t,\lambda)-I)\mathrm{d}\lambda
=&\d\int_{|\lambda-i\lambda_{0}|=\epsilon}((M^{a})_+^{-1}(\varsigma,t,\lambda)-I)\mathrm{d}\lambda  \vspace{0.15in} \\
&\d\quad+\int_{|\lambda-i\lambda_{0}|=\epsilon}
(\widetilde{\mu}(\varsigma,t,\lambda)-I)((M^{a})_+^{-1}(\varsigma,t,\lambda)-I)\mathrm{d}\lambda \vspace{0.15in}\\
=&\d -2\pi i\frac{(\delta^0_{i\lambda_0})^{\hat{\sigma}_3}M_{1}^Y(r(i\lambda_0))}{\frac{2\sqrt t}{\lambda^2_0}}+O\l(\frac{r(i\lambda_0)}{t}\r),\quad t\rightarrow\infty,\,
\end{array}
\ene
\bee \begin{array}{rl}\label{666}
\d\int_{|\lambda+i\lambda_{0}|=\epsilon}
\widetilde{\mu}(\varsigma,t,\lambda)((M^{a})_+^{-1}(\varsigma,t,\lambda)-I)\mathrm{d}\lambda
&=\d\int_{|\lambda+i\lambda_{0}|=\epsilon}((M^{a})_+^{-1}(\varsigma,t,\lambda)-I)\mathrm{d}\lambda  \vspace{0.15in} \\
&\d\quad +\int_{|\lambda+i\lambda_{0}|=\epsilon}
(\widetilde{\mu}(\varsigma,t,\lambda)-I)((M^{a})_+^{-1}(\varsigma,t,\lambda)-I)\mathrm{d}\lambda \vspace{0.15in}\\
&=\d -2\pi i\frac{(\delta^0_{-i\lambda_0})^{\hat{\sigma}_3}M_{1}^Y(r(-i\lambda_0))}{\frac{2\sqrt t}{\lambda^2_0}}+O\l(\frac{r(-i\lambda_0)}{t}\r),\quad t\rightarrow\infty.
\end{array}
\ene

On the other hand, we have
\bee\begin{array}{rl}
\d\left|\int_{\Sigma'}
\widetilde{\mu}(\varsigma,t,\lambda)\widetilde{w}(\varsigma,t,\lambda)\mathrm{d}\lambda\right|
&=\d \left|\int_{\Sigma'}
(\widetilde{\mu}(\varsigma,t,\lambda)-I)\widetilde{w}(\varsigma,t,\lambda)\mathrm{d}\lambda+\int_{\Sigma'}
\widetilde{w}(\varsigma,t,\lambda)\mathrm{d}\lambda\right| \vspace{0.1in}\\
&\leq \d\|\widetilde{\mu}-I\|_{L^2(\Sigma')}\|\widetilde{w}\|_{L^2(\Sigma')}+\|\widetilde{w}\|_{L^1(\Sigma')}.
\end{array}
\ene

Nowadays, according to Eq.~(\ref{5}) and Proposition 5.1, we obtain
\bee\begin{array}{l}
\d\left|\int_{\Sigma'}
\widetilde{\mu}(\varsigma,t,\lambda)\widetilde{w}(\varsigma,t,\lambda)\mathrm{d}\lambda\right|=O(Ct^{-\frac{3}{2}}), \v\\
\d\left|\int_{\Sigma^{(4)}\setminus\Sigma'}
\widetilde{\mu}(\varsigma,t,\lambda)\widetilde{w}(\varsigma,t,\lambda)\mathrm{d}\lambda\right|=O\l(\frac{\ln t}{t}\r).
\end{array}\ene
Since
\bee\begin{array}{rl}
\lim\limits_{\lambda\rightarrow\infty}\lambda(M^{(4)}(\varsigma,t,\lambda)-I)
=&\!\!\!\d\lim\limits_{\lambda\rightarrow\infty}\lambda(\widetilde{M}(\varsigma,t,\lambda)-I)=
\frac{i}{2\pi}\int_{\widetilde{\Sigma}}\widetilde{\mu}(\varsigma,t,\lambda)\widetilde{w}(\varsigma,t,\lambda)\mathrm{d}\lambda \vspace{0.1in}\\
=&\d\frac{i}{2\pi}\int_{|\lambda-\lambda_{0}|=\epsilon}\widetilde{\mu}(\varsigma,t,\lambda)\widetilde{w}(\varsigma,t,\lambda)\mathrm{d}\lambda
+\frac{i}{2\pi}\int_{|\lambda+\lambda_{0}|=\epsilon}\widetilde{\mu}(\varsigma,t,\lambda)\widetilde{w}(\varsigma,t,\lambda)\mathrm{d}\lambda\v\\
&\d+\frac{i}{2\pi}\int_{|\lambda-i\lambda_{0}|=\epsilon}\widetilde{\mu}(\varsigma,t,\lambda)\widetilde{w}(\varsigma,t,\lambda)\mathrm{d}\lambda
+\frac{i}{2\pi}\int_{|\lambda+i\lambda_{0}|=\epsilon}\widetilde{\mu}(\varsigma,t,\lambda)\widetilde{w}(\varsigma,t,\lambda)\mathrm{d}\lambda \v\\
&\d+\frac{i}{2\pi}\int_{\Sigma'}
\widetilde{\mu}(\varsigma,t,\lambda)\widetilde{w}(\varsigma,t,\lambda)\mathrm{d}\lambda \d+\frac{i}{2\pi }\int_{\Sigma^{(4)}\setminus\Sigma'}
\widetilde{\mu}(\varsigma,t,\lambda)\widetilde{w}(\varsigma,t,\lambda)\mathrm{d}\lambda\vspace{0.1in}\\
=&\!\!\!\d\frac{i}{2\pi}\int_{|\lambda-\lambda_{0}|=\epsilon}
\widetilde{\mu}(\varsigma,t,\lambda)((M^{a})_+^{-1}(\varsigma,t,\lambda)-I)\mathrm{d}\lambda\v\\
&\d+\frac{i}{2\pi}\int_{|\lambda+\lambda_{0}|=\epsilon}
\widetilde{\mu}(\varsigma,t,\lambda)((M^{a})_+^{-1}(\varsigma,t,\lambda)-I)\mathrm{d}\lambda\v\\
&\d+\frac{i}{2\pi}\int_{|\lambda-i\lambda_{0}|=\epsilon}
\widetilde{\mu}(\varsigma,t,\lambda)((M^{a})_+^{-1}(\varsigma,t,\lambda)-I)\mathrm{d}\lambda\v\\
&\d+\frac{i}{2\pi }\int_{|\lambda+i\lambda_{0}|=\epsilon}
\widetilde{\mu}(\varsigma,t,\lambda)((M^{a})_+^{-1}(\varsigma,t,\lambda)-I)\mathrm{d}\lambda\v\\
&\d+\frac{i}{2\pi }\int_{\Sigma'}
\widetilde{\mu}(\varsigma,t,\lambda)\widetilde{w}(\varsigma,t,\lambda)\mathrm{d}\lambda
\d+\frac{i}{2\pi }\int_{\Sigma^{(4)}\setminus\Sigma'}
\widetilde{\mu}(\varsigma,t,\lambda)\widetilde{w}(\varsigma,t,\lambda)\mathrm{d}\lambda.
\end{array}
\ene

Therefore, it follows from Eq.~(\ref{666}) that we have
\bee\begin{array}{rl}
\lim\limits_{\lambda\rightarrow\infty}\lambda(\widetilde{M}(\varsigma,t,\lambda)-I)
=&\!\!\!\d-\frac{\lambda^2_0(\delta^0_{\lambda_0})^{\hat{\sigma}_3}M_{1}^X(r(\lambda_0))}{2\sqrt t}-\frac{\lambda^2_0(\delta^0_{-\lambda_0})^{\hat{\sigma}_3}M_{1}^X(r(-\lambda_0))}{2\sqrt t} \v\\
&\d+\frac{\lambda^2_0(\delta^0_{i\lambda_0})^{\hat{\sigma}_3}M_{1}^Y(r(i\lambda_0))}{2\sqrt t}+\frac{\lambda^2_0(\delta^0_{-i\lambda_0})^{\hat{\sigma}_3}M_{1}^Y(r(-i\lambda_0))}{2\sqrt t}
+O\l(\frac{\ln t}{t}\r),\quad t\rightarrow\infty.
\end{array}
\ene
which further leads to
\bee\begin{array}{rl}
\lim\limits_{\lambda\rightarrow\infty}(\lambda\widetilde{M}(\varsigma,t,\lambda))_{12}
=&\!\!\!\d-\frac{i\lambda_0^2\beta^{X}(r(\lambda_0))(\delta_{\lambda_0}^0)^2}{2\sqrt t}-\frac{i\lambda_0^2\beta^{X}(r(-\lambda_0))(\delta_{-\lambda_0}^0)^2}{2\sqrt t} \v\\
&\d+\frac{i\lambda_0^2\beta^{Y}(r(i\lambda_0))(\delta_{i\lambda_0}^0)^2}{2\sqrt t}
+\frac{i\lambda_0^2\beta^{Y}(r(-i\lambda_0))(\delta_{-i\lambda_0}^0)^2}{2\sqrt t}
+O\l(\frac{\ln t}{t}\r),\quad t\rightarrow\infty.
\end{array}
\ene

It follows from the symmetry reduction for $M^X$ and $M^Y$~\cite{xu15} that we have
\bee\begin{array}{l}
\beta^{X}(r(\lambda_0))=\beta^{X}(r(-\lambda_0)), \v\\
\beta^{Y}(r(i\lambda_0))=\beta^{Y}(r(-i\lambda_0)),
\end{array}
\ene
such that we find
\bee\label{mg}\begin{array}{rl}
\lim\limits_{\lambda\rightarrow\infty}(\lambda\widetilde{M}(\varsigma,t,\lambda))_{12}
=& \d-\frac{i\lambda_0^2\beta^{X}(r(\lambda_0))[(\delta_{\lambda_0}^0)^2+
(\delta_{-\lambda_0}^0)^2]}{2\sqrt t} \v\\
&\d +\frac{i\lambda_0^2\beta^{Y}(r(i\lambda_0))\{(\delta_{i\lambda_0}^0)^2+
(\delta_{-i\lambda_0}^0)^2\}}{2\sqrt t}+O\l(\frac{\ln t}{t}\r),\quad t\rightarrow\infty.
\end{array}
\ene

Based on the properties of Ref.~\cite{lenells2}, that is
\bee\no
a(-\lambda)=a(\lambda),\,\, b(-\lambda)=-b(\lambda),\,\, A(-\lambda)=A(\lambda),\,\, B(-\lambda)=-B(\lambda),
\ene
we have
\bee\no\begin{array}{l}
r(-z)=-r(z), \v\\
\chi_{\pm}(\lambda_0)=\chi_{\mp}(-\lambda_0), \quad
\widetilde{\chi'}_{\pm}(\lambda_0)=\widetilde{\chi'}_{\mp}(-\lambda_0), \v\\
\chi'_{\pm}(i\lambda_0)=\chi_{\mp}(-i\lambda_0),\quad
\widetilde{\chi}_{\pm}(i\lambda_0)=\widetilde{\chi}_{\mp}(-i\lambda_0).
\end{array}
\ene

Thus it follows from the above-mentioned equations that we have
\bee \label{mp}\begin{array}{rl}
\lim\limits_{\lambda\rightarrow\infty}(\lambda\widetilde{M}(\varsigma,t,\lambda))_{12}
=&\!\!\!\!\!\d\frac{-i\lambda_0^2\sqrt\upsilon}{\sqrt t}e^{i\left[\!\!\frac{\pi}{4}+\arg r(\lambda_0)-\arg\Gamma(-i\upsilon(r(\lambda_0)))+
2\tilde{\upsilon}\ln(2\lambda_0^2)-\upsilon\ln\!\frac{\lambda_0^2}{t}
+2t-\frac{t}{\lambda_0^2}+2i\chi_{\pm}(\lambda_0)+2i\widetilde{\chi'}_{\pm}(\lambda_0)\right]} \v\\
&\!\!\!\!\!+\d\frac{i\lambda_0^2\sqrt{\tilde{\upsilon}}\emph{}}{\sqrt t}e^{i\left[\!\!\frac{\pi}{4}+\arg r(i\lambda_0)+\arg\Gamma(i\tilde{\upsilon}(r(i\lambda_0)))-
2\upsilon\ln(2\lambda_0^2)+\tilde{\upsilon}\ln\!\frac{\lambda_0^2}{t}
+2t+\frac{t}{\lambda_0^2} +2i\chi'_{\pm}(i\lambda_0)+2i\widetilde{\chi}_{\pm}(i\lambda_0)\right]}\v\\
&\!\!\!\!\!\d+O\l(\frac{\ln t}{t}\r).
\end{array}
\ene

\v
\noindent {\it 5.2. \, Long-time asymptotics of $q_x(x,t)$} \v

According to Eq.~(\ref{mp}) and
\bee
 m(x,t)=\lim\limits_{\lambda\rightarrow\infty}(\lambda M(x,t,\lambda))_{12}
=\lim\limits_{\lambda\rightarrow\infty}(\lambda\widetilde{M}(\varsigma,t,\lambda))_{12},\quad t\to \infty,
\ene
we have the following properties.

\v\noindent {\bf Proposition 5.4} As $t\to \infty$
\bee
\begin{array}{rl}
m(x,t)=&\!\!\!\lim\limits_{\lambda\rightarrow\infty}(\lambda M)_{12}=\lim\limits_{\lambda\rightarrow\infty}(\lambda\widetilde{M}(\varsigma,t,\lambda))_{12}\v\\
=&\!\!\!\d\frac{-i\lambda_0^2\sqrt\upsilon}{\sqrt t}e^{i(\frac{\pi}{4}+\arg r(\lambda_0)-\arg\Gamma(-i\upsilon(r(\lambda_0))+
2\tilde{\upsilon}\ln{2\lambda_0^2}-\upsilon\ln\frac{\lambda_0^2}{t}
+2t-\frac{t}{\lambda_0^2}+2i\chi_{\pm}(\lambda_0)+2i\widetilde{\chi'}_{\pm}(\lambda_0))} \v\\
&+\d\frac{i\lambda_0^2\sqrt{\tilde{\upsilon}}}{\sqrt t}e^{i(\frac{\pi}{4}+\arg r(i\lambda_0)+\arg\Gamma(i\tilde{\upsilon}(r(i\lambda_0))-
2\upsilon\ln{2\lambda_0^2}+\tilde{\upsilon}\ln\frac{\lambda_0^2}{t}
+2t+\frac{t}{\lambda_0^2} +2i\chi'_{\pm}(i\lambda_0)+2i\widetilde{\chi}_{\pm}(i\lambda_0))} \v\\
&\d +O\l(\frac{\ln t}{t}\r).
\end{array}
\ene

\noindent {\bf Proposition 5.5} \, As $t\to \infty$, we have
\bee\begin{array}{rl}
\d\int_0^x 2|m(x',t)|^2\mathrm{d}x'=&\!\!\! \d 2\int_0^x [(\Re m(x',t))^2+(\Im m(x',t))^2]dx' \v\\
=&\!\!\! \d 2\int_0^x\l[\frac{\lambda'^4\upsilon+\lambda'^4|\tilde{\upsilon}|}{t}+
\frac{\lambda'^4\sqrt{\upsilon|\tilde{\upsilon}|}}{t}\sin(\eta_2-\eta_1)\r]dx'+
O(x\frac{\ln t}{t^{\frac{3}{2}}})+O\l(x\frac{(\ln t)^2}{t^2}\r)\v\\
=&\!\!\!\d -2\int_0^{\lambda_0}\l[\frac{\upsilon(\lambda')+|\tilde{\upsilon}(\lambda')|}{\lambda'}+
\frac{\sqrt{\upsilon|\tilde{\upsilon}|(\lambda')}}{\lambda'}\sin(\eta_2(\lambda')-\eta_1(\lambda'))\r]d\lambda'+
O\l(\frac{\ln t}{t^{\frac{1}{2}}}\r)
\end{array}
\ene
where $(x'=\frac{t}{4\lambda'^4}-t)$ and
\bee\no\begin{array}{c}
\eta_1(\lambda)=\d\frac{\pi}{4}+\arg r(\lambda)-\arg\Gamma(-i\upsilon(r(\lambda)))+
2\tilde{\upsilon}\ln{2\lambda^2}-\upsilon\ln\frac{\lambda^2}{t}
+(2-\lambda^{-2})t+2i[\chi_{\pm}(\lambda)+\widetilde{\chi'}_{\pm}(\lambda)],\v\\
\eta_2(\lambda)=\d\frac{\pi}{4}+\arg r(i\lambda)+\arg\Gamma(i\tilde{\upsilon}(r(i\lambda)))-
2\upsilon\ln{2\lambda^2}+\tilde{\upsilon}\ln\frac{\lambda^2}{t}
+(2+\lambda^{-2})t +2i[\chi'_{\pm}(i\lambda)+\widetilde{\chi}_{\pm}(i\lambda)].
\end{array}
\ene

Since
\bee\label{sp}
q_x(x,t)= 2im(x,t)e^{2i\int_{(0,0)}^{(x,t)}\Delta},
\ene
and
\bee
\Delta(x,t)=\frac{1}{2}|q_x|^2dx+\frac{1}{2}(|q_x|^2-|q|^2)dt.
\ene
thus we have
\bee\label{delta}
\Delta=2|m|^2dx-2\l(\int_x^\infty (|m|^2)_t \mathrm{d}x'\r)dt=2|m|^2dx+\frac{1}{2}(|q_x|^2-|q|^2)dt
\ene

Therefore, according to Eq.~(\ref{delta}) and boundary-value conditions, by choosing the special integral contour for $\Delta$,  we have

\v\noindent {\bf Proposition 5.6}
\bee\begin{array}{rl}
\d\int_{(0,0)}^{(x,t)}\Delta=&\d\int_{(0,0)}^{(0,t)}\Delta+\int_{(0,t)}^{(x,t)}\Delta \v\\
=&\d\frac{1}{2}\int_0^t(|g_1|^2-|g_0|^2)\mathrm{d}t'+\int_0^x 2|m(x',t)|^2\mathrm{d}x'
\end{array}\ene

Therefore, According to Propositions 5.4, 5.5, 5.6, and Eq.~(\ref{sp}), we can show that Theorem 1.1 holds.

\v\noindent
{\bf Acknowledgments} \vspace{0.05in}

This work was partially supported by NSFC under Grant No.11571346 and the Youth Innovation Promotion Association CAS.


\end{document}